\documentclass[12pt]{article}
\usepackage{cite,epsfig,amssymb,amsmath,graphicx,color}
\evensidemargin \oddsidemargin

\newcommand{\nn}{\nonumber}

\begin{document}

\begin{center}
{{{\Large \bf Mass-deformed ABJM Theory and LLM Geometries: Exact Holography
}
}\\[17mm]
Dongmin Jang$^{1}$,~~Yoonbai Kim$^{1}$,
~~O-Kab Kwon$^{1}$,~~D.~D. Tolla$^{1,2}$\\[3mm]
{\it $^{1}$Department of Physics,~BK21 Physics Research Division,
~Institute of Basic Science, Sungkyunkwan University, Suwon 440-746, South Korea\\
$^{2}$University College,\\
Sungkyunkwan University, Suwon 440-746, South Korea}\\[2mm]
{\it dongmin@skku.edu,~yoonbai@skku.edu,~okab@skku.edu,~ddtolla@skku.edu} }

\end{center}
\vspace{15mm}

\begin{abstract}
We present a detailed account and extension of our claim in arXiv:1610.01490. 
We test the gauge/gravity duality between the ${\cal N} = 6$ mass-deformed ABJM theory with  U$_k(N)\times$U$_{-k}(N)$ gauge symmetry and the 11-dimensional supergravity  on LLM geometries with SO(4)/${\mathbb Z}_k$ $\times$SO(4)/${\mathbb Z}_k$ isometry, 
in the large $N$ limit.   
Our analysis is based on the evaluation of vacuum expectation values of chiral primary operators from the supersymmetric vacua of  mass-deformed ABJM theory and from the implementation of Kaluza-Klein holography to the LLM geometries. 
We focus on the chiral primary operator with conformal dimension $\Delta = 1$.  We show that $\langle {\cal O}^{(\Delta=1)}\rangle= N^{\frac32} \, f_{(\Delta=1)}$ for all supersymmetric vacuum solutions and LLM geometries with  $k=1$, where the factor $f_{(\Delta)}$ is independent of $N$. We  also confirm that the vacuum expectation value of the the energy momentum tensor is vanishing as expected by the supersymmetry. We extend our results to the case of  $k\ne 1$ for LLM geometries represented by rectangular-shaped Young-diagrams. In analogy with the Coulomb branch of the ${\cal N} = 4$ super Yang-Mills theory, we argue that the discrete Higgs vacua of the mABJM theory as well as the corresponding LLM geometries are parametrized by the vevs of the chiral primary operators.  
\end{abstract}

\newpage
\tableofcontents

\section{Introduction}

Gauge/gravity duality conjecture states an equivalence between a theory of quantum gravity in $(d+1)$-dimensional spacetime and a quantum field theory (QFT) on the $d$-dimensional boundary of the spacetime~\cite{Maldacena:1997re,Gubser:1998bc,Witten:1998qj}.
In its original context~\cite{Maldacena:1997re}, the duality was conjectured for string/M theory on AdS$_{d+1}\times {\cal X}$, with a compact internal manifold ${\cal X}$, and conformal field theory (CFT) living on $d$-dimensional boundary of the AdS space.
Having the application to realistic theories like QCD in mind, the extension of the conjecture to non-conformal field theories was pursued soon after~\cite{Klebanov:1999tb}.
In particular, the conjecture was extended to QFTs that are obtained from the CFTs either by adding relevant operators to the action or considering vacua where the conformal symmetries are broken spontaneously.
A $d$-dimensional QFT, which is obtained as a result of either of those deformations, is dual to a string/M theory on a spacetime geometry which is asymptotically AdS$_{d+1}\times {\cal X}$.
However, there is no complete formulation of string/M theory on a curved background.
Hence, the duality is mainly tested in the limit of a weakly curved classical gravity, which corresponds to taking the limits of large $N$ as well as large 't Hooft coupling constant $\lambda$, $N$ being the rank of the gauge group.

One of the tests of gauge/gravity duality involves the calculation of the vacuum expectation values (vevs) of gauge invariant operators in the large $N$ and $\lambda$ limits. On the field theory side, the calculation follows the usual perturbation expansion where the divergences in the bare quantities are subtracted using the standard renormalization procedure.
When those gauge invariant operators are chiral primary operators (CPOs) in highly supersymmetric gauge theories, the vevs are protected from quantum corrections by the supersymmetry and they are determined in the classical limit.
The corresponding procedure on the gravity side goes as follows.
Given a supergravity solution which is asymptotically AdS$_{d+1}\times {\cal X}$, it is expanded in terms of harmonic functions on the compact manifold.
The compactification results in towers of Kaluza-Klein (KK) modes in $(d+1)$-dimensional gravity theory.
The gauge/gravity correspondence dictates that for every gauge invariant operator on the field theory side there is a corresponding gravity field among these KK modes.
The vev of the gauge invariant operator is then determined by applying the holographic renormalization procedure~\cite{Henningson:1998gx} to those KK modes in the $(d+1)$-dimensional gravity theory~\cite{Skenderis:2006uy,Skenderis:2006di,Skenderis:2007yb}.
For a CPO of conformal dimension $\Delta$, the vev which is obtained using this procedure is proportional to the coefficient of $z^\Delta$ in the asymptotic expansion of the dual scalar field, $z$ being the holographic coordinate of the AdS space.
This procedure was implemented to determine the vevs of CPOs in the Coulomb branch of the ${\cal N}=4$ super Yang-Mills (SYM) theory and the dual type IIB string theory on a spacetime geometry which is asymptotically AdS$_{5}\times {S^5}$ \cite{Skenderis:2006uy,Skenderis:2006di}.

In \cite{Jang:2016tbk}, we reported a summary of our work which shows an exact gauge/gravity duality relation for large $N$. 
Our analysis is based on the 3-dimensional mass-deformed Aharony-Bergman-Jafferis-Maldacena theory (mABJM) of massive M2-branes, which has ${\cal N}=6$ supersymmetry and ${\rm U}_{k}(N)\times {\rm U}_{-k}(N)$ gauge symmetry, where $k$ is the Chern-Simons level~\cite{Hosomichi:2008jb,Gomis:2008vc}.
The mass-deformed theory is obtained from the original ABJM theory~\cite{Aharony:2008ug} by adding a relevant deformation which preserves the full supersymmetry as well as the gauge symmetry while the conformal symmetry is completely broken and the SU$(4)$ global symmetry is reduced to SU$(2)\times $SU$(2)\times$U$(1)$.
The mABJM theory supports sets of discrete Higgs vacua, which are expressed in terms of the GRVV matrices with numerical valued matrix elements~\cite{Gomis:2008vc}.
The calculation of the vevs of CPOs in the large $N$ limit is possible due to the existence of these discrete vacua.

Since the mABJM theory is obtained from the deformation of a CFT by relevant operators, the spacetime geometry of the dual gravity theory should be asymptotically AdS$_4\times{\cal X}$.
It was predicted that for the gauge theory describing massive M2-branes, the dual gravity theory is M-theory on the 11-dimensional Lin-Lunin-Maldacena (LLM) geometry~\cite{Bena:2004jw, Lin:2004nb}.
Indeed, the LLM geometry with ${\mathbb Z}_k$ orbifold and SO(2,1)$\times$SO(4)/${\mathbb Z}_k\times$SO(4)/${\mathbb Z}_k$ isometry is asymptotically AdS$_4\times S^7/{\mathbb Z}_k$.
In line with this prediction a one-to-one correspondence between the vacua of the mABJM theory and the LLM geometry was obtained~\cite{Kim:2010mr,Cheon:2011gv}. See also \cite{Hashimoto:2011nn,Kwon:2011nv,Massai:2014wba} for related works.

In this paper, we quantitatively test the above gauge/gravity duality in terms of the vevs of CPOs. 
On the field theory side, some of the classical vacuum solutions are protected from quantum corrections due to the high number of supersymmetry.
The vevs of CPOs are determined by those supersymmetric vacua.
We calculate the vevs of the CPO with conformal dimension $\Delta=1$ for all possible supersymmetric vacua of the mABJM theory for large $N$  and general $k$.
On the gravity theory side, we start with equations of motion on AdS$_4\times S^7$ background in 11-dimensional supergravity.
In order to obtain the 4-dimensional equations of motion, we implement the KK reduction procedure on $S^7$.
The LLM solutions were obtained in some unknown gauge.
In order to solve the 4-dimensional equations using the LLM solutions, we need to write the equations with gauge invariant combinations for KK modes\footnote{For the linearized equations of motion on AdS$_4\times S^7$ background in de Donder gauge, see \cite{Biran:1983iy}.
It is important to note that, recovering the equations of motion of gauge invariant fields from those of the fields in the de Donder gauge is not straightforward.}.
In general, the KK reduction leads to cubic or higher order interaction terms among the KK modes, which results in non-linear equations of motion.
When we are interested only in the CPO with conformal dimension $\Delta=1$, the linearized equations are sufficient.
However, for CPOs with $\Delta \ge 2$, one has to consider non-linear equations\cite{JKKT}, where some non-trivial field redefinitions are required to relate 4-dimensional and 11-dimensional fields.
See \cite{Lee:1998bxa,Arutyunov:1999en,Lee:1999pj, Skenderis:2006uy} for non-linear results on the AdS$_5\times S^5$ background.
Here we focus on the $\Delta=1$ case.
According to the gauge/gravity dictionary~\cite{Gubser:1998bc,Witten:1998qj}, we read the vevs from the asymptotic expansions of the KK scalar modes.
As a result, we obtain an exact holographic relation which is given by
\begin{align}
\langle {\cal O}^{(\Delta=1)}\rangle= N^{\frac32} \, f_{(\Delta=1)},
\end{align}
where we consider the $k=1$ case, $f_{(\Delta)}$ is a function of the conformal dimensions and also depends on some parameters of the LLM solutions, but does not depend on $N$.
For a given $N$ the number of supersymmetric vacua is equal to the partition of $N$ and the above result is valid for all supersymmetric vacua~\cite{Kim:2010mr,Cheon:2011gv}.
We also extend this result to $k > 1$, however, for some specific types of the LLM solutions.

The paper is organized as follows.
In section \ref{mABJMLLM}, we summarize the one-to-one correspondence between the discrete supersymmetric vacua of the mABJM theory and the LLM solutions with ${\mathbb Z}_k$ orbifold.
In section \ref{VCPO-mABJM}, we discuss CPOs in the ABJM theory and obtain the vevs of the operators in the case $\Delta=1$.
In section \ref{KKR-GIV}, we apply the KK reduction to 11-dimensional supergravity equations of motion and obtain linearized equations for 4-dimensional gauge invariant KK modes.
In section \ref{KKH-LLM}, we use the method of holographic renormalization to read the vevs of CPO with $\Delta=1$ from the asymptotic expansion of the LLM solutions and compare the field theory and the gravity results.
In section \ref{CONC}, we draw some conclusions and discuss some future directions as well. We also include three appendixes where we discuss some general features of spherical harmonics on $S^7$, give some details about asymptotic expansions of the LLM solutions as well as the proof of equation \eqref{Identity}.

\section{Vacua of the mABJM Theory and the LLM Geometries}\label{mABJMLLM}

The ${\cal N}=6$ ABJM theory with ${\rm U}(N)_k\times {\rm U}(N)_{-k}$ gauge group is a superconformal CS matter theory with CS level $k$ and it describes the low energy dynamics of $N$ coincident M2-branes on the ${\mathbb C}^4/ {\mathbb Z}_k$ orbifold fixed point~\cite{Aharony:2008ug}.
One interesting feature of the ABJM theory is that it allows supersymmetry preserving mass deformation~\cite{Hosomichi:2008jb,Gomis:2008vc}.
That is, the resulting mass-deformed theory called the mABJM theory has still ${\cal N}=6$ supersymmetry though the conformal symmetry of the original theory is broken under the deformation.
This deformation is achieved by adding some terms to the Lagrangian of the ABJM theory, which break the global SU(4) symmetry to ${\rm SU(2)}\times {\rm SU(2)}\times {\rm U}(1)$.
Solving the classical vacuum equation of the mABJM theory, it was shown that the classical vacuum solutions are discrete and represented by the GRVV matrices~\cite{Gomis:2008vc}.
Some vacuum solutions for given $N$ and $k$ are protected from quantum corrections and have one-to-one correspondence with the LLM geometries with ${\mathbb Z}_k$~\cite{Kim:2010mr,Cheon:2011gv} quotient in 11-dimensional supergravity.
In this section, we briefly review the correspondence.

\subsection{Supersymmetric vacua of the mABJM theory}\label{Vac-mABJM}
To reflect the global symmetry of the mABJM theory we split the 4-complex scalar fields as follows
\begin{align}
Y^A=(Z^a,W^{\dagger a}),
\end{align}
where $A=1,2,3,4$ and $a,b=1,2$.
Then the vacuum equation of the mABJM theory, ${\cal L}_{{\rm bos}}=-V_{{\rm bos}}=0$, is written as
\begin{align}\label{Vac-eq}
&Z^aZ^\dagger_bZ^b-Z^bZ^\dagger_bZ^a=-\frac{\mu k}{2\pi} Z^a,\qquad W^{\dagger a}W_b W^{\dagger b}
-W^{\dagger b} W_b W^{\dagger a}=\frac{\mu k}{2\pi} W^{\dagger a},\nn\\
&W_aZ^bW_b-W_bZ^bW_a=0,\qquad Z^bW_bZ^a-Z^aW_bZ^b=0,
\end{align}
where $\mu$ is a mass parameter.
The general solutions of the matrix equations in \eqref{Vac-eq} have been found in the form of the GRVV matrices~\cite{Gomis:2008vc}.
For given $N$ and $k$, there are many possible solutions satisfying the equations in \eqref{Vac-eq}.
A systematic way to classify the vacuum solutions is to represent those as direct sums of two types of irreducible $n\times (n+1)$ matrices, ${\cal M}_a^{(n)}~(a=1,2)$ and their Hermitian conjugates, $\bar{\cal M}_a^{(n)}$.
These rectangular matrices are the GRVV matrices
\begin{align}\label{mat-1}
{\cal M}_1^{(n)}&=\left(\begin{array}{cccccc}
\sqrt{n\!}\!\!\!&0&&&&\\&\!\sqrt{n\!-\!1} \!\!&\!0&&&\\
&&\ddots&\ddots&&\\&&&\sqrt{2}&0&\\&&&&1&0\end{array}\right),
\qquad
{\cal M}_2^{(n)}&=
\left(\begin{array}{cccccc}0&1&&&&\\&0&\sqrt{2}&&&\\ &&\ddots&\ddots&&\\
&&&0\!&\!\!\sqrt{n\!-\!1}\!&\\&&&&0&\!\!\!\sqrt{n\!}\end{array}
\right),
\end{align}
where $n = 0, 1,\cdots, N-1$.
The vacuum solutions are given by
\begin{align}\label{ZW-vacua}
Z^a_0&=\sqrt{\frac{\mu k}{2\pi}}\left(\begin{array}{c}
\begin{array}{cccccc}\mathcal{M}_a^{(n_1)}\!\!&&&&&\\&\!\!\ddots\!&&&&\\
&&\!\!\mathcal{M}_a^{(n_i)}&&& \\ &&& {\bf 0}_{(n_{i+1}+1)\times n_{i+1}}
&&\\&&&&\ddots&\\&&&&&{\bf 0}_{(n_f+1)\times n_f}\end{array}\\
\end{array}\right),\nonumber
\end{align}
\begin{align}
W^{\dagger a}_0&=\sqrt{\frac{\mu k}{2\pi}}\left(\begin{array}{c}
\begin{array}{cccccc}{\bf 0}_{n_1\times (n_1+1)}&&&&&\\&\ddots&&&&\\
&&{\bf 0}_{n_i\times(n_i + 1)} &&&\\
&&& \bar{\mathcal M}_a^{(n_{i+1})}\!\!&&\\&&&&\!\!\ddots\!&\\
&&&&&\!\!\bar{\mathcal M}_a^{(n_f)}\end{array}\\
\end{array}\right).
\end{align}
The solution contains $N_n$ rectangular matrices of the type ${\cal M}_a^{(n)}$ and $N_n'$ rectangular matrices of the type $\bar{\cal M}_a^{(n)}$.
From now on we refer to $N_n$ and $N_n'$ as occupation numbers~\cite{Kim:2010mr,Cheon:2011gv}.
Here $N_0$ and $N_0'$ denote the numbers of empty columns and rows, respectively.
Since $Z^a$ and $W^{\dagger a}$ are $N\times N$ matrices, the occupation numbers, $N_n$ and $N_n'$, should satisfy the following two constraints,
\begin{align}\label{NnNnp}
N = \sum_{n=0}^{N-1}\Big[\left(n+\frac12\right)\left(N_n+ N_n'\right)\Big], 
\qquad 
 \sum_{n=0}^{\infty}N_n = \sum_{n=0}^{\infty}N_n'. 
\end{align} 
At quantum level, only a subset of these classical solutions, which satisfy the conditions,
\begin{align}\label{susyvacuum}
0\le N_n\le k,\qquad 0\le N_n'\le k,
\end{align}
remains to be supersymmetric~\cite{Kim:2010mr}.

\subsection{LLM geometries and their droplet picture}\label{LLMgeom}
The LLM solution with SO(2,1)$\times$SO(4)$\times$SO(4) isometry in 11-dimensional supergravity is conjectured to be dual to the theory of massive M2-branes~\cite{Bena:2004jw,Lin:2004nb}.
Later, the mABJM theory with CS level $k=1$ is proposed to be the theory of massive M2-branes.
For the mABJM theory with general $k$, one has to consider the $\mathbb{Z}_k$ orbifold of the LLM geometry as the dual gravity theory~\cite{Cheon:2011gv}.

The LLM geometry with $\mathbb{Z}_k$ orbifold is given by
\begin{align}\label{LLMmet}
ds^2 = -{\bf G}_{tt} ( -dt^2 + dw_1^2 + dw_2^2) + {\bf G}_{xx} (d\tilde x^2 + d\tilde y^2) 
+ {\bf G}_{\theta\theta} ds^2_{S^3/\mathbb{Z}_k}
+ {\bf G}_{\tilde\theta\tilde\theta} ds^2_{\tilde S^3/\mathbb{Z}_k},
\end{align}
where $ds^2_{S^3/\mathbb{Z}_k}$ and $ds^2_{\tilde S^3/\mathbb{Z}_k}$ are metrics of the two $S^3$'s with $\mathbb{Z}_k$ orbifold and the warp factors are given by
\begin{align}\label{warpfac}
&{\bf G}_{tt}= -\left(\frac{4\mu_0^2 \tilde y\sqrt{\frac14 - Z^2}}{f^2}
\right)^{2/3},\qquad 
{\bf G}_{xx}=\left(\frac{f\sqrt{\frac14 - Z^2}}{2 \mu_0y^2}\right)^{2/3},
\nn \\
&{\bf G}_{\theta\theta} =\left(
\frac{f\tilde y \sqrt{\frac12 + Z}}{2 \mu_0\left(\frac12 -Z\right)} \right)^{2/3},
\qquad
{\bf G}_{\tilde\theta\tilde\theta}=\left( \frac{f\tilde y \sqrt{\frac12 - Z}}{2\mu_0 
\left(\frac12 + Z\right)}\right)^{2/3}
\end{align}
with
\begin{align}
f(\tilde x,\tilde y) = \sqrt{1 - 4 Z^2 - 4\tilde y^2 V^2}, \qquad \mu_0=\frac{\mu}4.
\end{align}
As we see in \eqref{warpfac}, the geometry is completely determined by two functions, which are given by
\begin{align}\label{ZV}
Z(\tilde x,\tilde y)
=\sum_{i=1}^{2N_B\!+\!1}\frac{(-1)^{i\!+\!1}
(\tilde x\!-\!\tilde x_i)}{2\sqrt{(\tilde x\!-\!\tilde x_i)^2+\tilde y^2}}
\ ,\qquad
V(\tilde x,\tilde y)
=\sum_{i=1}^{2N_B\!+\!1}\frac{(-1)^{i\!+\!1}}{2\sqrt{(\tilde x\!-\!\tilde x_i)^2+\tilde y^2}},
\end{align}
where $\tilde x_i$'s are the positions of the boundaries between the black and the white regions and $N_B$ is the number of finite size black regions in the droplet representation, as we will see below.
We also note that the two functions satisfy the relation, $\tilde ydV=-\star_2\ dZ~{\rm with}~\epsilon_{\tilde y\tilde x}=1$.
The corresponding 4-form field strength is given by
\begin{align}\label{LLMF4}
{\bf F}_4 &= -d \left(e^{2\Phi}h^{-2}V \right)
\wedge dt\wedge dw_1\wedge dw_2 +\mu_0^{-1} \left[Vd(\tilde y^2e^{2G}) + h^2e^{3G}\star_2 d(\tilde y^2 e^{-2G})\right]
\wedge d\Omega_3
\nn \\
&~~~+\mu_0^{-1}
\left[ Vd(\tilde y^2e^{-2G}) -h^2e^{-3G}\star_2 d(\tilde y^2 e^{2G})\right]
\wedge d\tilde\Omega_3,
\end{align}
where $d\Omega_3=-(\sin\theta/8)d\theta\wedge d\phi\wedge d\psi$, $d\tilde\Omega_3=-(\sin\tilde\theta/8)d\tilde\theta\wedge d\tilde\phi\wedge d\tilde\psi$ in the Euler coordinate system.\footnote{Vielbeins for $S^3$ in terms of the Euler angles are given by $\sigma^1=-\sin\psi d\theta+\sin\theta\cos\psi d\phi,\,\,\sigma^2=\cos\psi d\theta+\sin\theta\sin\psi d\phi,\,\,\sigma^3=d\psi+\cos\theta d\phi$ with ranges of the angles, $ 0\le \theta \le \pi, \,\, 0\le \phi \le 2\pi,\,\, 0\le\psi \le 4\pi$.
The metric on the 3-sphere and the volume form with unit radius are written as $ds_{S^3}^2=\frac14\left(d\theta^2 +\sin^2\theta d\phi^2+\left(d\psi+\cos\theta d\phi\right)^2\right)$.}
The $\mathbb{Z}_k$ quotient acts as $\left(\psi, \, \tilde\psi\right) \to \left( \psi + \frac{4\pi}{k},\, \tilde\psi + \frac{4\pi}{k}\right)$\cite{Cheon:2011gv,Auzzi:2009es}.
The 4-form field strength in \eqref{LLMF4} can also be expressed in terms of $Z(\tilde x,\tilde y)$ and $V(\tilde x,\tilde y)$ by using the relations~\cite{Kim:2014yca,Kim:2016dzw}
\begin{align}
h^2
=
\frac{\sqrt{\frac14 - Z^2}}{\tilde y},
\qquad 
e^{2\Phi}
=
\frac{4 \tilde y \mu_0^2 \sqrt{\frac14 - Z^2}}{f^2},
\qquad 
e^{2 G}
=
\frac{\frac12 + Z}{\frac12 - Z}.
\end{align}

We note that the function $Z(\tilde x,\tilde y)$ at $\tilde y=0$ has a value $\frac12$ if $\tilde x_{2i-1}< \tilde x < \tilde x_{2i}$ and it has a value $-\frac12$ if $\tilde x_{2i}< \tilde x < \tilde x_{2i+1}$.
Based on this fact, the LLM geometries are represented in terms of an infinite strip in the $\tilde x$-direction with regions of $Z(\tilde x,0)=-\frac12$ denoted by black color and regions of $Z(\tilde x,0)=\frac12$ denoted by white color.
This is called the droplet representation.
See the Fig.1.
Since the function $Z(\tilde x,0)$ is $ -\frac12$ if $\tilde x<\tilde x_1$ and it is $\frac12$ if $\tilde x>\tilde x_{2N_B+1}$, $N_B$ being the number of finite black regions, we note that the strip also contains an infinite black region below $\tilde x_1$ and an infinite white region above $\tilde x_{2N_B+1}$.

For every droplet picture there is a symmetric point such that the length of all finite size black regions above this point is the same as the length of all finite size white regions below the point.
This point is called the Fermi level $\tilde x_F$, which is given by
\begin{align}\label{xF}
\tilde x_F=\tilde x_1+\sum_{i=1}^{N_B}(\tilde x_{2i+1}-\tilde x_{2i}).
\end{align}
The strip is divided into excitation levels above and below the Fermi level, where each level has length $k$.
The levels are labeled by non-negative integers $n=0,1,2,\cdots$ starting at the Fermi level.
A given droplet representation is then parametrized by a set of parameters $\{l_n,l_n'\}$ with $l_n$ corresponding to the length of the black region in the $n$-th level above the Fermi level and $l_n'$ corresponding to the length of the white region in the $n$-th level below the Fermi level.
Since the length of the black or white region in a given level cannot be bigger than $k$ these parameters should satisfy the condition $0\le l_n,l_n'\le k$, which is the same as \eqref{susyvacuum}.
Actually, it have been suggested that there is one-to-one correspondence between the LLM solutions and the vacua of mABJM theory~\cite{Cheon:2011gv}.
Since, the LLM solutions are classified by $\{l_n, \, l_n'\}$, while the field theory vacua are classified by the occupation numbers $\{N_n,\, N_n'\}$, the one-to-one correspondence is given by
\begin{align}\label{NNn}
\{l_n,\, l_n'\}
\Longleftrightarrow
\{N_n,\, N_n'\}. 
\end{align}

An alternative representation of the LLM solutions is given in terms of Young diagrams.
In Young diagram representation, the lengths of the white and black regions correspond to the lengths of the horizontal and vertical edges of the Young diagram, respectively.
See Fig.1 for the parametrization of droplet picture and Young diagram.\\ 
\begin{figure}
\centerline{\epsfig{figure=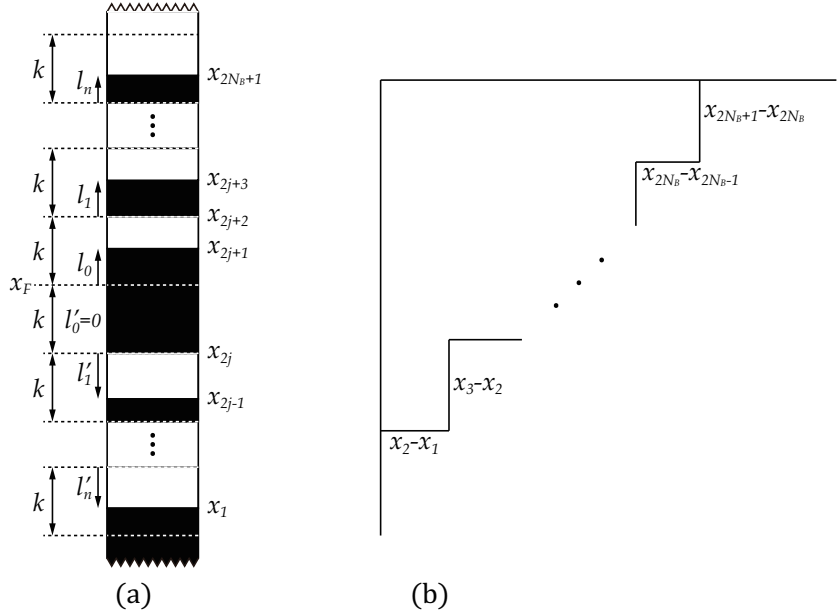,height=90mm}}
\caption{
\small (a) A droplet representation of the LLM geometry with SO(2,1)$\times$SO(4)/${\mathbb Z}_k\times$SO(4)/${\mathbb Z}_k$ isometry. The horizontal width does not correspond to any coordinate but added for clarity. (b) The Young diagram corresponding to the droplet picture (a).}
\label{Fig1}
\end{figure}

\section{Vevs of CPOs in mABJM Theory}\label{VCPO-mABJM}
For the Coulomb branch of the ${\cal N}=4$ SYM theory, the adjoint scalar fields $X^i$'s satisfy the vacuum equation $[X^i, \, X^j] = 0$, which means the matrix representations of the scalar fields describing the vacuum moduli are diagonal.
As a result, the U($N$) gauge symmetry is broken to ${\rm U}(1)^N$.
The vacuum moduli preserve the ${\cal N}=4$ supersymmetry while the conformal symmetry is completely broken.
The vevs of CPOs are non-renormalizable due to high supersymmetry and they can parametrize the Coulomb branch vacua.
On the other hand, in the type IIB supergravity, some BPS solutions describing D3-branes distributed over finite region of the transverse space were obtained~\cite{Kraus:1998hv}.
These solutions are asymptotic to AdS$_5\times S^5$.
According to the gauge/gravity dictionary, the vevs of CPOs is read from the asymptotic expansion of the dual scalar fields.
Calculating the vevs of CPOs with lower conformal dimensions, 
the exact dual relations for the Coulomb branch of the ${\cal N} =4$ SYM theory in the large $N$ limit were tested in a systematic way~\cite{Skenderis:2006uy,Skenderis:2006di}.

Now we compare the mABJM theory and the Coulomb branch of the 
SYM theory.
There are some differences between these two theories.
For instance, the mABJM theory is constructed by adding some relevant terms to Lagrangian of the ABJM theory and have discrete Higgs vacua with matrix representations composed of numerical elements, while for the Coulomb branch of the ${\cal N} =4$ SYM theory, the Lagrangian is undeformed, instead it is defined by choosing some non-vanishing vacuum moduli, which are composed of continuous parameters.
However, these two theories are similar in the sense that the theories are away from the UV fixed point and they preserve the full supersymmetry with the dual geometries asymptotic to AdS times a compact manifold.
Based on these facts and the known results of the Coulomb branch of the ${\cal N} =4$ SYM theory, one can expect that similar phenomena may happen for the Higgs vacua in the mABJM theory.
That is, the Higgs vacua of the mABJM theory are parametrized by vevs of CPOs and those vevs are read from asymptotic expansions of the LLM geometries using the holographic renormalization procedure.

In this section, we construct the CPO with $\Delta = 1$, which manifests the global symmetry of the mABJM theory.
We also calculate the vevs of the oprator for all supersymmetric vacua of mABJM theory for general $k$ in the large $N$ limit. 

\subsection{CPOs in ABJM theory}\label{CPOs-def}
The gauge invariant CPOs of conformal dimension $\Delta$ in the ABJM theory are given by
\begin{align}\label{CPO}
{\cal O}^{(\Delta)}=C_{A_1,\cdots,A_{n}}^{(\Delta)B_1,\cdots,B_{n}}{\rm Tr}\big(Y^{A_1}Y^\dagger_{B_1}\cdots Y^{A_{n}}Y^\dagger_{B_{n}}\big),
\end{align}
where $A,B\cdots = 1,\cdots, 4$ and $C_{A_1,\cdots,A_{n}}^{(\Delta)B_1,\cdots,B_{n}}$ are symmetric in lower as well as upper indices and traceless when tracing over one lower index and one upper index.
The CPO in \eqref{CPO} is written by manifesting the global SU(4) symmetry of the ABJM theory.
On the other hand, in the mABJM theory the CPOs have to manifest the SU(2)$\times$SU(2)$\times$U(1) global symmetry.

It is well known fact that the coefficients $C_{A_1,\cdots,A_{n}}^{(\Delta)B_1,\cdots,B_{n}}$ are identified with the similar coefficients $C^{I}_{i_1\cdots i_{2n}}$, which defines the scalar spherical harmonics on $S^7$ (see appendix \ref{SHS7})~\cite{Klebanov:1999tb}.
These coefficients also satisfy the same orthonormality condition as those of the spherical harmonics;
\begin{align}
C_{A_1,\cdots,A_{n}}^{(\Delta_1)B_1,\cdots,B_{n}}\bar C_{B_1,\cdots,B_{n}}^{(\Delta_2)A_1,\cdots,A_{n}}+ (\textrm{c.c.})= \delta^{\Delta_1\Delta_2}.
\end{align}
Therefore, one can fix these coefficients knowing the corresponding coefficients of the spherical harmonics on $S^7$. In appendix \ref{Scalar-Harm}, we list the first few scalar spherical harmonics on $S^7$, which are needed to read the coefficients for CPOs with lower conformal dimensions.
In particular, the coefficients of the CPO with $\Delta=1$ are determined in appendix \ref{CabD1} and the operator is given by
\begin{align}\label{Delta1}
{\cal O}^{(\Delta=1)}=\frac{1}{2\sqrt{2}}{\rm Tr}\big(Z^{a}Z^\dagger_{a}-W^{\dagger a}W_{a}\big).
\end{align}

\subsection {vevs of CPO in mABJM Theory} \label{vev-CPO}
Here we calculate the vevs of CPO with $\Delta=1$ in mABJM theory.
As we see in \eqref{CPO}, the CPO we are considering is composed of the complex scalar fields $Y^A$'s and their complex conjugates.
For a given supersymmetric vacuum, the scalar fields near the vacuum are expanded as
\begin{align}\label{Yexp}
Y^A = Y_0^A + \hat Y^A,
\end{align}
where $Y_0^A$'s ($A=1,2,3,4$) denote the vacuum solutions represented by the GRVV matrices~\cite{Gomis:2008vc}, and $\hat Y^A$'s are field operators.
Inserting \eqref{Yexp} into \eqref{CPO} of a CPO with conformal dimension $\Delta$, we obtain
\begin{align}\label{vevcal}
\langle {\cal O}^{(\Delta)}\rangle_m = {\cal O}^{(\Delta)}(Y^A_0) 
+ \sum_i \langle \delta {\cal O}_i^{(\Delta)}\rangle_0 +\frac1N{\rm -corrections},  
\end{align}
where $\langle  \cdots \rangle_m$ and  $\langle  \cdots \rangle_0$ denote the vevs of operators in the mABJM theory and the ABJM theory, respectively, and $\delta {\cal O}_i^{(\Delta)} $ is an operator containing at least one $\hat Y^A$ or $\hat Y^{\dagger A}$. 
The $\frac1N$-corrections in \eqref{vevcal} come from the contributions of multi-trace terms~\cite{Witten:2001ua,Berkooz:2002ug,Gubser:2002vv}.  Here we also note that quantum corrections of scalar fields are absent due to the high number of supersymmetry of the mABJM theory.  The second term in \eqref{vevcal} is a one point function in a conformal field theory and is vanishing. Therefore, in the large $N$ limit we have
\begin{align}\label{vevCPO1}
\langle {\cal O}^{(\Delta)}(Y^A)\rangle_{m} &= {\cal O}^{(\Delta)}(Y_0^A).
\end{align}

For CPO with $\Delta=1$ the vevs are obtained by plugging the vacuum solutions in \eqref{ZW-vacua} into \eqref{Delta1}.
Since $Z^a_0$ is a block diagonal matrix, which contains $N_n$ of the rectangular matrix ${\cal M}_a^{(n)}$, while $W^{\dagger a}_0$ contains $N'_n$ of the matrix $\bar{\cal M}_a^{(n)}$, we have
\begin{align}
{\rm Tr}(Z^a_0Z^\dagger_{a0}-W^{\dagger a}_0W_{a0})=\frac{\mu k}{2\pi}\sum_{n=0}^{\infty}(N_n-N_n'){\rm Tr}({\cal M}_a^{(n)}\bar{\cal M}_a^{(n)})=\frac{\mu k}{2\pi}\sum_{n=0}^{\infty}(N_n-N_n')n(n+1),
\end{align}
where in the last step we have used the rectangular matrices in \eqref{mat-1}.
Then, we obtain
\begin{align}\label{vevD1}
\langle {\cal O}^{(\Delta=1)}\rangle_m=\frac{\mu k}{4\sqrt2\pi}\sum_{n=0}^{\infty}(N_n-N_n')n(n+1).
\end{align}
This result is valid for all supersymmetric vacua of mABJM theory with finite $k$ in the large $N$ limit.

\section {KK Reduction and Gauge Invariant Modes}\label{KKR-GIV}
The KK reduction of 11-dimensional gravity to 4 dimensions involves compactification of the fields on $S^7$.\footnote{The ABJM theory is dual to M-theory on AdS$_4\times S^7/{\mathbb Z}_k$. Here we consider $k=1$ case, for simplicity. We will extend our results to general $k$ case eventually.}
In this section we apply the KK reduction to the 11-dimensional gravity on AdS$_4\times S^7$ to obtain 4-dimensional equation of motion on AdS$_4$ background.
Such KK reduction was carried out in the de Donder gauge in \cite{Biran:1983iy}.
However, the LLM solutions of our interest in this paper are in a different gauge and can not be analyzed based on the results obtained in the de Donder gauge.
Therefore, we carry out the reduction in a generic gauge and write the equation of motion for gauge invariant dynamical fields.

\subsection{11-dimensional gravity equations of motion}
The bosonic part of the 11-dimensional supergravity action is given by
\begin{align}\label{act1}
S&= \frac{1}{16\pi G_{11} }\int d^{11}x \Big[\sqrt{-{\bf g}}\big({\bf R}-\frac1{48}{\bf F}_{pqrs}{\bf F}^{pqrs}\big)+ \frac{1}{2(4!)^2}\tilde\epsilon^{p_1p_2p_3q_1\cdots
q_4r_1\cdots r_4} {\bf C}_{p_1p_2p_3}{\bf F}_{q_1\cdots q_4} {\bf F}_{r_1\cdots
r_4}\Big],
\end{align}
where we used the index notation $p,q,r,\cdots =0,\cdots,10$, ~$\tilde\epsilon^{0123\cdots 10}=-1$ is the Levi-Civita symbol.
The 11-dimensional Newton's gravitational constant is
\begin{align}
 G_{11} = \frac{1}{32\pi^2} (2\pi l_{\rm P})^9,
\end{align}
where $l_{\rm P}$ is the Planck constant.
The functional variation of the action gives the following equations of motion for the metric and the 4-form field strength:
\begin{align}\label{11dEOM}
&{\bf R}_{pq}-\frac12{\bf g}_{pq}{\bf R}=\frac{1}{48}
\Big(-\frac12{\bf g}_{pq}{\bf F}_{rstu}{\bf F}^{rstu}+4{\bf F}_{pstu}{\bf F}_q^{~stu}\Big),\nn
\\
&\partial_p({\bf e}{\bf F}^{pqrs})+\frac{1}{2\cdot (4!)^2}\tilde\epsilon^{p_1\cdots p_4q_1\cdots q_4qrs}{\bf F}_{p_1\cdots p_4}{\bf F}_{q_1\cdots q_4}=0,
\end{align}
where ${\bf e}\equiv \sqrt{-{\bf g}}$.
Using the index notation $(\mu,\nu,\rho,\cdots =0,\cdots 3),~ (a,b,c,\cdots=4,\cdots 10)$ we write the AdS$_4\times S^7$ solution of the equations of motion in \eqref{11dEOM} as follows
\begin{align}\label{AdS4S7}
&ds^2=\frac{L^2}{ 4\rho^2}\left(-dt^2+dw_1^2+ dw_2^2+d\rho^2\right)+ L^2 ds^{2}_{S^{7}},\nn\\
&F_{\mu\nu\rho\sigma}=-\frac6{L}\epsilon_{\mu\nu\rho\sigma}, ~{\rm and~it~is~zero~otherwise}. 
\end{align}
Here $\epsilon_{\mu\nu\rho\sigma} = \sqrt{|g_{{\rm AdS_4}}|}
\,\tilde \epsilon_{\mu\nu\rho\sigma}$ is the Levi-Civita tensor for the AdS$_4$ space, and $L$ is the radius of $S^7$.

\subsection{Fluctuations on AdS$_4\times S^7$}\label{AstEOM}
We consider a solution which is asymptotically AdS$_4\times S^7$ so that we can write it as
\begin{align}\label{flucgC}
{\bf g}_{pq}&=g_{pq}+ h_{pq},
\nn \\
{\bf C}_{pqr}&=C_{pqr}+ c_{pqr}\quad
\Longleftrightarrow
\quad 
{\bf F}_{pqrs}=F_{pqrs}+ f_{pqrs},
\end{align}
where $h_{pq}, c_{pqr}, f_{pqrs}$ represent deviations from the AdS$_4\times S^7$ geometry and they become small fluctuations in the asymptotic region\footnote{We use a notation in which the objects in 11-dimensional supergravity are denoted by bold font symbols whereas the AdS$_4\times S^7$ values of those objects are denoted by normal font symbols.}.
Plugging this back into \eqref{11dEOM}, we obtain the following equations of motion for $h_{pq}~{\rm and}~ f_{pqrs}$ up to linear order,
\begin{align}
&\nabla^r\nabla_{p}h_{qr}+\nabla^r\nabla_{q}h_{pr}-\nabla^2h_{pq}-\nabla_q\nabla_ph^r_{~r}-Rh_{pq}-g_{pq}\left(-R^{rs}h_{rs}
+\nabla^r\nabla^sh_{rs}-\nabla^2h^r_{~r}\right)
\nn\\
&+\frac{1}{48}\Big(F_{rstu}F^{rstu} h_{pq}
{-4}g_{pq}h_{rs}F^r_{~tuv}F^{stuv}\Big)+\frac{1}{24}g_{pq}
f_{pqrs}F^{pqrs}\nn\\
&-\frac12 h_{rs}F^r_{~ptu}F_q^{~stu}-\frac1{6}\Big(
f_{prst}F_q^{~rst}+F_{prst}f_q^{~rst}\Big)=0,
\label{lineq2-1}
\end{align}
\begin{align}
&\partial_p(e f^{pqrs})+\frac12\partial_p(e h^p_pF^{pqrs})
+4\partial_p(e h_t^{[p}F^{qrs]t})
+\frac{1}{(4!)^2}\tilde\epsilon^{p_1\cdots p_4q_1\cdots q_4qrs} 
f_{p_1\cdots p_4}F_{q_1\cdots q_4}=0.
\label{lineq2-2}
\end{align}
It is convenient to write the above equations for the $AdS_4$ and $S^7$ indices separately
\begin{align}\label{lineq2-5}
&\nabla^\rho\nabla_{\mu}h_{\nu\rho}+\nabla^\rho\nabla_{\nu}h_{\mu\rho}
+\nabla^a\nabla_{\mu}h_{\nu a}+\nabla^a\nabla_{\nu}h_{\mu a}-(\nabla^\rho\nabla_\rho+\nabla^a\nabla_a) h_{\mu\nu}-\nabla_\mu\nabla_\nu (h^\rho_{~\rho}+h^a_{~a})+\frac6{L^2}h_{\mu\nu}\nn\\
&-g_{\mu\nu}\left[\frac{12}{L^2}h^\rho_{~\rho}-\frac6{L^2}h^a_{~a}+\nabla^\rho\nabla^\sigma h_{\rho\sigma}+\nabla^a\nabla^b h_{ab}+(\nabla^\rho\nabla^a+\nabla^a\nabla^\rho) h_{\rho a}-(\nabla^\rho\nabla_\rho+\nabla^a\nabla_a) (h^\sigma_{~\sigma}+h^b_{~b})\right]\nn\\
&+\frac{1}{48}\Big(F_{\rho\sigma\lambda\kappa}F^{\rho\sigma\lambda\kappa}h_{\mu\nu}{-
4}g_{\mu\nu}h_{\rho\sigma}F^\rho_{~\tau\lambda\kappa}F^{\sigma \tau\lambda\kappa}\Big)+\frac1{24}g_{\mu\nu}
f_{\rho\sigma\tau\lambda}F^{\rho\sigma\tau\lambda}{-\frac12}h_{\rho\sigma}F^\rho_{~\mu \tau\lambda}F_\nu^{~\sigma \tau\lambda}\nn\\
&-\frac1{6}\Big(f_{\mu \rho\sigma\tau}F_\nu^{~\rho\sigma\tau}+f_{\nu \rho\sigma\tau}F_\mu^{~\rho\sigma\tau}\Big)=0,
\end{align}
\begin{align}\label{lineq2-6}
&\nabla^\rho\nabla_{\mu}h_{a\rho}+\nabla^\rho\nabla_{a}h_{\mu\rho}
+\nabla^b\nabla_{\mu}h_{a b}+\nabla^b\nabla_{a}h_{\mu b}-(\nabla^\rho\nabla_\rho+\nabla^b\nabla_b) h_{\mu a}-\nabla_\mu\nabla_a (h^\rho_{~\rho}+h^b_{~b})+\frac6{L^2}h_{\mu a}\nn\\
&+\frac{1}{48}F_{\nu\rho\sigma\tau}F^{\nu\rho\sigma\tau}h_{\mu a}-\frac1{6}f_{a \rho\sigma\tau}F_\mu^{~\rho\sigma\tau}=0,
\end{align}
\begin{align}\label{lineq2-7}
&\nabla^\rho\nabla_{a}h_{b\rho}+\nabla^\rho\nabla_{b}h_{a\rho}
+\nabla^c\nabla_{a}h_{b c}+\nabla^c\nabla_{b}h_{a c}-(\nabla^\rho\nabla_\rho+\nabla^c\nabla_c) h_{ab}-\nabla_a\nabla_b (h^\rho_{~\rho}+h^c_{~c})+\frac6{L^2}h_{ab}\nn\\
&-g_{ab}\left[\frac{12}{L^2}h^\rho_{~\rho}-\frac6{L^2}h^c_{~c}+\nabla^\rho\nabla^\sigma h_{\rho\sigma}+\nabla^c\nabla^d h_{cd}+(\nabla^\rho\nabla^c+\nabla^c\nabla^\rho) h_{\rho c}-(\nabla^\rho\nabla_\rho+\nabla^c\nabla_c) (h^\sigma_{~\sigma}+h^d_{~d})\right]\nn\\
&+\frac{1}{48}\Big(F_{\mu\nu\rho\sigma}F^{\mu\nu\rho\sigma}h_{ab}{-
4}g_{ab}h_{\rho\sigma}F^\rho_{~\tau\lambda\kappa}F^{\sigma \tau\lambda\kappa}\Big)+\frac1{24}g_{ab}f_{\rho\sigma\tau\lambda}F^{\rho\sigma\tau\lambda}=0,
\end{align}
and
\begin{align}
&\nabla_\sigma f^{\sigma\mu\nu\rho}+\nabla_a f^{a\mu\nu\rho}
+\frac12(\nabla_\sigma h^\lambda_{~\lambda})F^{\sigma\mu\nu\rho}+\frac12 h^\lambda_{~\lambda}\nabla_\sigma F^{\sigma\mu\nu\rho}\nn\\
&+\frac12(\nabla_\sigma h^a_{~a})F^{\sigma\mu\nu\rho}+\frac12h^a_{~a}\nabla_\sigma F^{\sigma\mu\nu\rho}+4\nabla_\lambda\big(h_\sigma^{[\lambda}F^{\mu\nu\rho]\sigma}\big)
+\nabla_a\big(h_\sigma^{a}F^{\mu\nu\rho\sigma}\big)=0,
\label{C3EoM1-2}\\
&\nabla_\sigma f^{\sigma\mu\nu a}+\nabla_b f^{b\mu\nu a}-\nabla_\lambda(h_\sigma^{~a}F^{\mu\nu \lambda\sigma})=0,\label{C3EoM2-2}\\
&\nabla_\sigma f^{\sigma\mu ab}+\nabla_c f^{c\mu ab}
=0,\label{C3EoM3-2}\\
&\nabla_\sigma f^{\sigma abc}+\nabla_d f^{dabc}
+\frac{1}{(4!)^2}\epsilon^{a_1\cdots a_4\nu_1\cdots \nu_4 abc}f_{a_1\cdots a_4}F_{\nu_1\cdots \nu_4}=0,\label{C3EoM4-2}
\end{align}
where we have used the identities $\nabla_p f^{pqrs}=\frac 1e\partial_p\big(ef^{pqrs}\big)$ and $\nabla_p\big(h_t^{[p}F^{qrs]t}\big)=\frac1e \partial_p(eh_t^{[p}F^{qrs]t})$.

\subsection{Expansion in $S^7$ spherical harmonics}\label{eomSVT}
The fluctuations $h_{pq}$ and $c_{pqr}$ can be expanded in $S^7$ spherical harmonics as
\begin{align}\label{hpqexp}
&
h_{\mu\nu}(x,y)
=
h^{I_1}_{\mu\nu}(x)Y^{I_1}(y),
\nn\\
&h_{\mu a}(x,y)
=
v_\mu^{I_7}(x)Y_a^{I_7}(y)
+s^{I_1}_\mu(x)\nabla_{a}Y^{I_1}(y),
\nn\\
&
h_{(ab)}(x,y)
=
t^{I_{27}}(x)Y_{(ab)}^{I_{27}}(y)
+v^{I_7}(x)\nabla_{(a}Y_{b)}^{I_7}(y)
+s^{I_1}(x)\nabla_{(a}\nabla_{b)}Y^{I_1}(y),
\nn\\
&
h^a_{~a}(x,y)
=
\phi^{I_1}(x)Y^{I_1}(y), 
\nn \\
&c_{\mu\nu\rho}(x,y)
=
\tilde s_{\mu\nu\rho}^{I_1}(x)Y^{I_1}(y),
\nn\\
&
c_{\mu\nu a}(x,y)
=
\tilde v_{\mu\nu}^{I_7}(x)Y_a^{I_7}(y)
+\tilde s^{I_1}_{\mu\nu}(x)\nabla_{a}Y^{I_1}(y),
\nn\\
&
c_{\mu ab}(x,y)
=
\tilde t_\mu^{I_{21}}(x)Y_{[ab]}^{I_{21}}(y)
+\tilde v_\mu^{I_7}(x)\nabla_{[a}Y_{b]}^{I_7}(y)
,
\nn\\
&
c_{abc}(x,y)
=
\tilde t^{I_{35}}(x)Y_{[abc]}^{I_{35}}(y)
+\tilde t^{I_{21}}(x)\nabla_{[a}Y_{bc]}^{I_{21}}(y),
\end{align}
where $x$ is the $AdS_4$ coordinate and $y$ is the $S^7$ coordinate.
For the definitions of the spherical harmonics on $S^7$, see appendix \ref{SHS7}. The notation $(ab)$ means symmetrized traceless combination which is defined as
\begin{align}
T_{(ab)}
=
\frac12(T_{ab}+T_{ba})
-\frac17g_{ab}T^c_{~c},
\end{align}
where $g_{ab}$ is a metric on $S^7$.
The notation $[abc\cdots]$ means anti-symmetrization among indices, $a,b,c,\cdots$, for instance,
\begin{align}
T_{[ab]}
=\frac1{2!}(T_{ab}-T_{ba}).
\end{align}
The expansion \eqref{hpqexp} follows the convention of \cite{Kim:1985ez,Skenderis:2006uy}.
The expansions of the 4-form field strength fluctuations $f_{pqrs}$ are read from $f=dc$,
\begin{align}\label{F4-exp}
&
f_{\mu\nu\rho\sigma}(x,y)
=
4\nabla_{[\mu}s_{\nu\rho\sigma]}^{I_1}(x)Y^{I_1}(y),
\nn\\
&
f_{\mu\nu\rho a}(x,y)
=
3\nabla_{[\mu}v_{\nu\rho]}^{I_7}(x)Y_a^{I_7}(y)
-s^{I_1}_{\mu\nu\rho}(x)\nabla_{a}Y^{I_1}(y),
\nn\\
&
f_{\mu\nu ab}(x,y)
=
2\nabla_{[\mu}t_{\nu]}^{I_{21}}(x)Y_{[ab]}^{I_{21}}(y)
+2v_{\mu\nu}^{I_7}(x)\nabla_{[a}Y_{b]}^{I_7}(y)
,\nn\\
&
f_{\mu abc}(x,y)
=
\nabla_\mu t^{I_{35}}(x)Y_{[abc]}^{I_{35}}(y)
-3 t_\mu^{I_{21}}(x)\nabla_{[a}Y_{bc]}^{I_{21}}(y)
,\nn\\
&
f_{abcd}(x,y)
=
4 t^{I_{35}}(x)\nabla_{[a}Y_{bcd]}^{I_{35}}(y),
\end{align}
where we have used the fact that $\nabla_{[a}\nabla_{b]}Y^{I_1}(y)=0$, $\nabla_{[a}\nabla_bY_{c]}^{I_7}(y)=0$, and $\nabla_{[a}\nabla_bY_{cd]}=0$.
We note that, under the $U(1)$ gauge transformation the 3-form gauge field transforms as $c_{3}\to d\Lambda_{(2)}$.
However, the 4-form field strength is invariant under this transformation.
Therefore, the field strengths in \eqref{F4-exp} are written in terms of $U(1)$ gauge invariant combinations which are defined as
\begin{align}\label{SVTT}
&
s^{I_1}_{\mu\nu\rho}(x)
\equiv
\tilde s^{I_1}_{\mu\nu\rho}(x)
-3\nabla_{[\mu}\tilde s_{\nu\rho]}^{I_1}(x),
\nn\\
&
v_{\mu\nu}^{I_7}(x)
\equiv
\tilde v_{\mu\nu}^{I_7}(x)
+\nabla_{[\mu}\tilde v_{\nu]}^{I_7}(x),
\nn\\
&
t_\mu^{I_{21}}(x)
\equiv
\tilde t_\mu^{I_{21}}(x)
-\frac{1}{3}\nabla_\mu \tilde t^{I_{21}}(x),
\nn\\
&
t^{I_{35}}(x)
\equiv
\tilde t^{I_{35}}(x).
\end{align}

Plugging \eqref{hpqexp} and \eqref{F4-exp} into the $(\mu,\, \nu)$-components of fluctuation equations in \eqref{lineq2-5} and projecting onto the scalar harmonics $ Y^{I_1}$, we obtain
\begin{align}\label{hmn_eq}
&
-
\Big(
\square
+\Lambda^{I_1}
+\frac{8}{L^2}
\Big)
h^{I_1}_{\mu\nu}
+\nabla_\mu\nabla^\rho h^{I_1}_{\nu\rho}
+\nabla_\nu\nabla^\rho h^{I_1}_{\mu\rho}
-g_{\mu\nu}\nabla^\rho\nabla^\sigma h^{I_1}_{\rho\sigma}
\nn\\
&
+g_{\mu\nu}
\Big(
\square
+\Lambda^{I_1}
-\frac{22}{L^2}
\Big)
h^{I_1}
-\nabla_\mu\nabla_\nu h^{I_1}
+\Lambda^{I_1}
\big(
\nabla_\mu s^{I_1}_\nu
+\nabla_\nu s^{I_1}_\mu
\big)
\nn\\
&
-2g_{\mu\nu}\Lambda^{I_1}\nabla^\rho s^{I_1}_\rho
+g_{\mu\nu}
\Big(
\square
+\frac67\Lambda^{I_1}
+\frac6{L^2}
\Big)
\phi^{I_1}
-\nabla_\mu\nabla_\nu \phi^{I_1}
-6g_{\mu\nu}\Lambda^{I_1}
\Big(
\frac17\Lambda^{I_1}
+\frac1{L^2}
\Big)
s^{I_1}
\nn\\
&
+\frac{4}{L}\nabla_{[\mu} s^{I_1}_{\rho\sigma\lambda]}\epsilon_\nu^{~\rho\sigma\lambda}
+\frac{4}{L} \nabla_{[\nu} s^{I_1}_{\rho\sigma\lambda]}
\epsilon_\mu^{~\rho\sigma\lambda}
-\frac{1}{L}g_{\mu\nu}\nabla_{[\rho} s^{I_1}_{\sigma\lambda\tau]}\epsilon^{\rho\sigma\lambda\tau}= 0,
\end{align}
where $\square\equiv\nabla_\mu\nabla^\mu$.
We have used the AdS$_4\times S^7$ solutions in \eqref{AdS4S7} and the results of the integrals of spherical harmonics listed in appendix \ref{SHS7}.
The tracing over $(\mu,\,\nu)$-indices in \eqref{hmn_eq} gives the following equation for scalar fields
\begin{align}\label{h-eqn}
&
\Big(
2\square
+3\Lambda^{I_1}
-\frac{96}{L^2}
\Big)
h^{I_1}
-2\nabla^\mu\nabla^\nu h_{\mu\nu}
-6\Lambda^{I_1}\nabla^\mu s^{I_1}_\mu
\nn\\
&
+3
\Big(
\square
+\frac87\Lambda^{I_1}
+\frac8{L^2}
\Big)
\phi^{I_1}
-24\Lambda^{I_1}
\Big(
\frac17\Lambda^{I_1}
+\frac1{L^2}
\Big)
s^{I_1}
+\frac4{L}\nabla_{[\mu} s^{I_1}_{\nu\rho\sigma]}\epsilon^{\mu\nu\rho\sigma}=0.
\end{align}

From the equation \eqref{lineq2-6}, we obtain the following two equations by projections onto $\nabla^a Y^{I_1}$ and $g^{ab}Y^{I_7}_b$, respectively
\begin{align}\label{leq2-2-2}
&
\Big(
\square
+\frac{24}{L^2}
\Big)
s_\mu^{I_1}
-\nabla_\mu\nabla^\nu s_\nu^{I_1}
-
\Big(
\frac67 \Lambda^{I_1}
+ \frac6{L^2}
\Big)
\nabla_\mu s^{I_1}
+\frac67\nabla_\mu\phi^{I_1}
\nn \\
&
-\nabla^\nu h_{\mu\nu}^{I_1}
+\nabla_\mu h^{I_1}
-\frac1{L} s^{I_1}_{\rho\sigma\tau} \epsilon_\mu^{~~\rho\sigma\tau}
=0,\\
-&\Big(\square +\Lambda^{I_7}+ \frac{18}{L^2}\Big)v_\mu^{I_7}
+ \nabla_\mu\nabla^\nu v_\nu^{I_7} + \Big(\frac12 \Lambda^{I_7}
+ \frac3{L^2}\Big) \nabla_\mu v^{I_7} {-\frac3L}\nabla_{[\nu}
v^{I_7}_{\lambda\kappa]}\epsilon_\mu^{~~\nu\lambda\kappa}=0.\label{leq2-2-2b}\end{align}
From the equations \eqref{lineq2-7}, we obtain four scalar equations by projecting on four different elements: $ g^{ab} Y^{I_1}$, $\nabla^{(a}\nabla^{b)} Y^{I_1}$, $\nabla^{(a} Y^{b)I_7}$ and $ Y^{(ab)I_{27}}$,
\begin{align}
-
&
\frac{12}{7}\Lambda^{I_1}\nabla^\mu s_\mu^{I_1}
-\frac1{L}\nabla_{[\mu}s_{\nu\lambda\kappa]}^{I_1}\epsilon^{\mu\nu\lambda\kappa}
-\frac{30}7\Lambda^{I_1}
\Big(
\frac{\Lambda^{I_1}}7
+\frac1{L^2}
\Big)
s^{I_1}
+
\Big(
\square
+\frac67\Lambda^{I_1}
+\frac{6}{L^2}
\Big)
h^{I_1}
\nn\\
&
+\frac67
\Big(
\square
+\frac{5}7\Lambda^{I_1}
+\frac{5}{L^2}
\Big)
\phi^{I_1}
-\nabla^\mu\nabla^\nu h_{\mu\nu}^{I_1}
=
0,\label{leq3-4-2}\\
&2\nabla^\mu s^{I_1}_\mu
=
\Big(
\square
-\frac57\Lambda^{I_1}
\Big)
s^{I_1}
+h^{I_1}
+\frac57\phi^{I_1}
,\label{leq3-3-2}\\
&\square v^{I_7}- 2\nabla^\mu v_\mu^{I_7}=- 2\nabla^\mu\hat{v}^{I_{7}}_{\mu}=0,
\label{leq3-3-3}\\
&\big( \square +\Lambda^{I_{27}} - \frac{2}{L^2}
\big)t^{I_{27}}=0.
\label{leq3-3-4}
\end{align}

Similarly, inserting \eqref{hpqexp} and \eqref{F4-exp} in to the equations of motion of $f_{pqrs}$ in \eqref{C3EoM1-2}-\eqref{C3EoM4-2}, and projecting onto the appropriate spherical harmonic elements, we obtain the following set of equations
\begin{align}
&
\Big(
\square
+\Lambda^{I_1}
\Big)
s^{I_1}_{\mu\nu\rho}
-3\nabla^{\delta}\nabla_{[\mu} s^{I_1}_{\nu\rho]\delta}
\nn\\
&
-\frac3{L}\epsilon_{\sigma\mu\nu\rho}\nabla^\sigma h^{I_1}
-\frac3{L}\epsilon_{\sigma\mu\nu\rho}\nabla^\sigma \phi^{I_1}
-\frac{24}{L}\nabla^\sigma h^{I_1}_{\lambda[\sigma}\epsilon_{\mu\nu\rho]}~\!\!^{\lambda}
-\frac6{L}\Lambda^{I_1}\epsilon_{\mu\nu\rho}~\!\!^{\sigma}s^{I_1}_\sigma=
0,
\label{tSmnr1} \\
&
\left(
\square
+\Lambda^{I_{7}}
+\frac{10}{L^{2}}
\right)
v^{I_{7}}_{\mu\nu}
-\frac{6}{L}\epsilon^{\rho}{}_{\sigma\mu\nu}\nabla^{\sigma}v^{I_{7}}_{\rho}=
0,
\label{tVmn7} \\
&
\left(
\square
+\Lambda^{I_{21}}
+\frac{2}{L^{2}}
\right)
t^{I_{21}}_{\mu}=
0,
\label{tT21}
\end{align}
\begin{align}
&
\nabla^{\rho} s^{I_1}_{\rho\mu\nu}
+\frac6{L}\epsilon_{\mu\nu\sigma}~\!\!^{\lambda}\nabla^\sigma s_\lambda^{I_1}=
\nabla^{\rho}\hat s^{I_1}_{\rho\mu\nu}
=
0, 
\label{tSmn1} \\
&
\nabla^{\nu} v^{I_{21}}_{\nu}=
0,
\label{tTm21}\\
&
\nabla^{\nu} v^{I_7}_{\nu\mu}=
0,
\label{tVm7}\\
&
\Big[
\square
-\frac{12}{L^2}
+\Lambda^{I_{35}}
-\frac{6\sqrt{\mu^{I_{35}}}}{L}
\Big]
\tilde t_+^{I_{35}}=
0, 
\nn\\
&
\Big[
\square
-\frac{12}{L^2}
+\Lambda^{I_{35}}
+\frac{6\sqrt{\mu^{I_{35}}}}{L}
\Big]
\tilde t_-^{I_{35}}=
0, 
\label{tT35}
\end{align}
where $\mu^{I_{35}}=\frac{(I_{35}+3)^2}{L^2}$ and we have used the relation
\begin{align}
\epsilon_{abc}~\!\!^{a_1a_2a_3a_4}\nabla_{a_1}Y^{I_{35}}_{a_2a_3a_4}
=\pm3!\sqrt{\mu^{I_{35}}}Y^{I_{35}}_{abc}.
\end{align}
to obtain the two equations in \eqref{tT35}.

\subsection{Gauge invariant fluctuations}
Some of the fluctuations $h_{pq}$ and $f_{pqrs}$ are related to each other or to the background solution by diffeomorphic transformation.
In other words, some of these fluctuations are generated by the variation of some other fluctuations or the background under the infinitesimal coordinate transformation $x'^p=x^p-\xi^p$.
Up to linear order, these gauge-dependent degrees of freedom transform as,
\begin{align}
& \tilde\delta h_{pq}=(\nabla_p\xi_q+\nabla_q\xi_p),
\qquad\tilde\delta f_{pqrs}=-4\nabla_{[p}\xi^tF_{qrs]t}.
\label{unphysical-dof}
\end{align}
To obtain the gauge transformations of the 4-dimensional fields, we expand the gauge parameter $\xi^p(x,y)$ in terms of the spherical harmonics on $S^7$ as
\begin{align}\label{gauge-par}
&\xi_\mu(x,y)=\xi_\mu^{I_1}(x)Y^{I_1}(y),\qquad\xi_a(x,y)=\xi_{(v)}^{I_7}(x)Y^{I_7}_a(y)+\xi^{I_1}_{(s)}(x)\nabla_aY^{I_1}(y).
\end{align}

Using \eqref{hpqexp}, \eqref{F4-exp}, and \eqref{gauge-par} in \eqref{unphysical-dof}, we obtain the following transformations for gauge-dependent coefficients of the spherical harmonics
\begin{align}
&\tilde \delta h_{\mu\nu}^{I_1}=\nabla_\mu\xi_\nu^{I_1}+\nabla_\nu\xi_\mu^{I_1}, \qquad \tilde \delta v_\mu^{I_7}=\nabla_\mu\xi_{(v)}^{I_7},\qquad \tilde \delta s_\mu^{I_1}=\nabla_\mu\xi_{(s)}^{I_1}+\xi_\mu^{I_1},\nn\\
&\tilde \delta t^{I_{27}}=0,\qquad \tilde \delta v^{I_7}=2\xi_{(v)}^{I_7},\qquad \tilde \delta s^{I_1}=2\xi_{(s)}^{I_1},
\qquad \tilde \delta\phi^{I_1}=2\Lambda^{I_1}\xi^{I_1}_{(s)}.
\end{align}
Based on these transformations, the following combinations are gauge invariant,
\begin{align}\label{ggi}
&\hat\phi^{I_1}=\phi^{I_1}-\Lambda^{I_1}s^{I_1},\qquad \hat v_\mu^{I_7}=v_\mu^{I_7}-\frac12\nabla_\mu v^{I_7},\qquad\hat h_{\mu\nu}^{I_1}=h_{\mu\nu}^{I_1}
-\nabla_{\mu}\tilde s_{\nu}^{I_1}
-\nabla_{\nu}\tilde s_{\mu}^{I_1},
\end{align}
where $\tilde s_\mu^{I_1}=s_\mu^{I_1}-\frac12\nabla_\mu s^{I_1}$.

Since the only non-vanishing component of the 4-form field strength is $F_{\mu\nu\rho\sigma}$, the non-trivial equations in second part of \eqref{unphysical-dof} are
\begin{align}
&\tilde \delta f_{\lambda\mu\nu\rho} =
-4 \nabla_{[\lambda}\left(\xi_\sigma^{I_1}
F_{\mu\nu\rho]}\!^\sigma \right)Y^{I_1},
\qquad\tilde \delta f_{\mu\nu\rho a} = \xi^{I_1}_\sigma F_{\mu\nu\rho}\!^{\sigma}\nabla_a Y^{I_1}.
\end{align}
Then, we obtain the following results
\begin{align}
\tilde \delta s_{\mu\nu\rho}^{I_1}=-\xi^{I_1}_\sigma F_{\mu\nu\rho}\!^\sigma,\qquad \nabla_{[\mu}\tilde \delta s_{\nu\rho]}^{I_1}
=\frac13 \left(\tilde\delta s_{\mu\nu\rho}^{I_1}
+ \xi^{I_1}_\sigma F_{\mu\nu\rho}\!^{\sigma}\right)=0,
\end{align}
The remaining coefficients associated with $f_{pqrs}$ are all diffeomorphic invariant at linear order.

\subsection{ KK reduction}\label{phys-modes}
The linear equations in subsection \ref{eomSVT} are not all independent and some of the fields are gauge degrees of freedom.
In this subsection, we find equations of motion for gauge invariant fields introduced perviously. We diagonalize those equations to identify the equations of motion for physical modes.

\subsubsection{The equations for spin zero fields}\label{Scalar-Eqs}
Using equations \eqref{h-eqn}, \eqref{leq2-2-2}, \eqref{leq3-4-2}, \eqref{leq3-3-2}, and \eqref{tSmnr1}, we obtain the following equations of motions for two gauge invariant scalar fields
\begin{align}
&\Big(\square+\Lambda^{I_1}+\frac{12}{L^2}
\Big)\hat\phi^{I_1}-\frac{14}{3L^2}\hat\psi^{I_1}=0,\label{h-eqn5} \\
&\Big(\square+\Lambda^{I_1}\Big)\hat\psi^{I_1}+18\Big(\square-\frac{5}{7}\Lambda^{I_1}\Big)\hat\phi^{I_1}=0\label{U-eqn5},
\end{align}
where the gauge invariant scalar fields are
\begin{align}\label{psiphi1}
&
\hat\psi^{I_1}
=
(18h^{I_1}- u^{I_1}),
\qquad
\hat\phi^{I_1}
=
(\phi^{I_1}-\Lambda^{I_1}s^{I_1})
\end{align}
with
\begin{align}\label{Utilde}
 u^{I_1}
\equiv
L\epsilon^{\mu\nu\rho\sigma}
\nabla_{\mu} s_{\nu\rho\sigma}^{I_1}.
\end{align}
Diagonalizing equations \eqref{h-eqn5} and \eqref{U-eqn5}, we write them in terms of the mass eigenstates
\begin{align}\label{diagpp}
\left(
\square
-\frac{({I_1}+6)({I_1}+12)}{L^2}
\right)
\check \phi^{I_1}
=
0,\qquad
\left(
\square
- \frac{{I_1}({I_1}-6)}{L^2}
\right)
\check\psi^{I_1}
=
0,
\end{align}
where we have introduced
\begin{align}\label{scme1}
\check\phi^{I_1}
=
\frac{({I_1}+7)\big[18({I_1}-1)\hat\phi^{I_1}+7\hat\psi^{I_1}\big]}{14({I_1}+3)},
\qquad
\check\psi^{I_1}
=
\frac{({I_1}-1)\big[-18({I_1}+7)\hat\phi^{I_1}+7\hat\psi^{I_1}\big]}{14({I_1}+3)}.
\end{align}

In addition, from equations \eqref{leq3-3-4} and \eqref{tT35} we write the equations of motion for three more fields which are already diagonal and gauge invariant
\begin{align}\label{Extra-Sca}
&\Big( \square -\frac{I_{27}(I_{27}+6)}{L^2}
\Big)\check t^{I_{27}}=0,\\
&\Big(
\square
-\frac{(I_{35}+3)(I_{35}+9)}{L^2}\Big)
\check t_+^{I_{35}}=
0, \qquad
\Big(
\square
-\frac{(I_{35}+3)(I_{35}-3)}{L^2}\Big)
\check t_-^{I_{35}}=
0,
\end{align}
where $\check t^{I_{27}}\equiv t^{I_{27}}$, $\check t_+^{I_{35}}\equiv t_+^{I_{35}}$, and $\check t_-^{I_{35}}\equiv t_-^{I_{35}}$.

In general, the KK reduction means to construct a 4-dimensional gravity action, including higher order interaction terms, from the equations of motion of fluctuation fields.
When one goes beyond the linear order, the higher order terms involve higher derivatives, and then one needs to introduce some field redefinitions in order to have the corresponding 4-dimensional gravity action.
For instance, at quadratic order, such field redefinition was introduced in \cite{Skenderis:2006uy,Skenderis:2006di}
\begin{align}\label{field-red1}
& S^{I} = \check s^{I}+J^s_{IJ_n J_m}\check t^{J_n}\check t^{J_m}
+L^s_{IJ_nJ_m}\nabla_\mu\check t^{J_n}\nabla^\mu\check t^{J_m},
\end{align}
where $\check s^{I}$ represent any of the 11-dimensional fields, and $S^{I}$ is the corresponding 4-dimensional field.
The $\check t^{J_i}$ represent all the fields that appear in the quadratic part of the equations of motion.
However, at linear order, which is what we are dealing with in this paper, these field redefinitions are trivial and the 11- and 4-dimensional fields are the same.
Therefore, the fields $\check \phi^{I_1},\check \psi^{I_1},\check t^{I_{27}},\check t_+^{I_{35}}, \check t_-^{I_{35}}$ are the correct 4-dimensional spin-zero fields at the linear order and are denoted as $ \Phi^{I_1}, \Psi^{I_1}, T^{I_{27}}, T_+^{I_{35}}, T_-^{I_{35}}$, respectively.
Based on the parity transformation of the original 11-dimensional fields ($h_{pq}$ is a tensor, $c_{pqr}$ is a pseudotensor), we note that the first three fields $\Phi^{I_1}, \Psi^{I_1}, T^{I_{27}}$ are scalar fields while the last two fields $T_+^{I_{35}}, T_-^{I_{35}}$ are pseudoscalar fields.

\subsubsection{The equations for spin one fields}\label{Vect-Eq}
There are three towers of KK vector modes. Combining equations \eqref{leq2-2-2b}, \eqref{leq3-3-3}, \eqref{tVmn7}, and \eqref{tVm7}, we obtain the following equations of motion for two of those KK modes
\begin{align}\label{vectEq}
&
\left(
\square
+\Lambda^{I_{7}}
+\frac{18}{L^{2}}
\right)
\hat{v}^{I_{7}}_{\mu}
+\frac{3}{L^{2}}\hat{u}^{I_{7}}_{\mu}=0,
\nn\\
&
\left(
\square
+\Lambda^{I_{7}}
+\frac{6}{L^{2}}
\right)
\hat{u}^{I_{7}}_{\mu}
+12
\left(
\square
+\frac{12}{L^{2}}
\right)
\hat{v}^{I_{7}}_{\mu}=0,
\end{align}
where $\Lambda^{I_{7}}=-\frac{I_7(I_7+6)-1}{L^{2}}$, and we have introduced the following gauge invariant combinations
\begin{align}
&\hat{v}^{I_{7}}_{\mu}
\equiv v^{I_{7}}_{\mu}
-\frac{1}{2}\nabla_{\mu}v^{I_{7}},\nn\\
&\hat{u}^{I_{7}}_{\mu}
\equiv
L\epsilon_{\mu}{}^{\nu\rho\sigma}\nabla_{\nu}\hat{v}^{I_{7}}_{\rho\sigma} \qquad{\rm with}~\hat{v}^{I_{7}}_{\mu\nu}
=\tilde{v}^{I_{7}}_{\mu\nu}
+\nabla_{[\mu}\tilde{v}^{I_{7}}_{\nu]}.
\end{align} 
Diagonalizing the two equations in \eqref{vectEq} gives the equations of motions for two mass eigenstates
\begin{align}
\left(
\square
-\frac{I_7^{2}+12I_7+23}{L^{2}}
\right)
\check{v}^{I_{7}}_{\mu}=0,
&
&
\left(
\square
-\frac{I_7^{2}-13}{L^{2}}
\right)
\check{u}^{I_{7}}_{\mu}=0, 
\end{align}
where
\begin{align}
\check{v}^{I_{7}}_{\mu}
=
\frac{2(I_7-1)\hat{v}^{I_{7}}_{\mu}-\hat{u}^{I_{7}}_{\mu}}{4(I_7+3)}
&,&
\check{u}^{I_{7}}_{\mu}
=
\frac{2(I_7+7)\hat{v}^{I_{7}}_{\mu}+\hat{u}^{I_{7}}_{\mu}}{4(I_7+3)}.
\end{align}

On the other hand from \eqref{tT21} we get the equation of motion for one more KK vector mode
\begin{align}
\left(\square+\Lambda^{I_{21}}+\frac2{L^2}\right)\check{t}_\mu^{I_{21}}
=0,
\end{align}
where $\check{t}_\mu^{I_{21}}$ is equivalent to $t_\mu^{I_{21}}$ in \eqref{SVTT}.

Like in the case of the scalar fields, if we consider non-linear equations, the KK reduction will involve non-trivial field redefinition of the type \eqref{field-red1}.
However, at linear order, the above three KK vector modes are the correct 4-dimensional physical modes and they are denoted as $V^{I_{7}}_{\mu}$, $U^{I_{7}}_{\mu}$, and $T_\mu^{I_{21}}$.
We note that the first two are vector fields while the third one is a pseudovector field.

\subsubsection{The equations for spin two fields}
Next we consider the gauge invariant equations of motion for 2-tensor fields.
Using equations \eqref{hmn_eq}, \eqref{leq2-2-2}, \eqref{leq3-4-2}, \eqref{leq3-3-2}, \eqref{tSmnr1}, and \eqref{tSmn1}, with some algebra, we obtain the following linear order equations for gauge invariant tensor fields
\begin{align}
&\Big(
\square
+\Lambda^{I_1}
+\frac{32}{L^2}
\Big)
\hat\psi_{\mu\nu}^{I_1}
-\frac{8}{L^2}g_{\mu\nu}\hat\psi^{I_1} 
+\frac{90}{7}
\Big(
\square
+\Lambda^{I_1}
+\frac{32}{L^2}
\Big)
\hat\phi_{\mu\nu}^{I_1}
-\frac{72}{7}
\Big(
3\nabla_{\mu}\nabla_{\nu}
+\frac{10}{L^2}g_{\mu\nu}
\Big)
\hat\phi^{I_1}
=0,
\nn \\
&\frac57
\Big(
\square
+\Lambda^{I_1}
+\frac{44}{L^2}
\Big)
\hat\phi_{\mu\nu}^{I_1}
+ \frac2{L^2}\hat\psi_{\mu\nu}^{I_1}
- \frac{40}{7L^2}g_{\mu\nu} \hat\phi^{I_1} 
- \frac{4}{3L^2} g_{\mu\nu}\hat\psi^{I_1}
=0,\label{phimn-eq}	
\end{align}
where we have introduced the following gauge invariant tensor fields
\begin{align}\label{psiphi2}
\hat\psi_{\mu\nu}^{I_1}
\equiv
18h^{I_1}_{\mu\nu}{-} u_{\mu\nu}^{I_1},
\qquad
\hat\phi_{\mu\nu}^{I_1}
\equiv
\frac75
\left(
\nabla_{\mu}s^{I_1}_{\nu}+\nabla_{\nu}s^{I_1}_{\mu}
-\nabla_{\mu}\nabla_{\nu}s^{I_1}
- h_{\mu\nu}^{I_1}
\right),
\end{align}
and we have defined
\begin{align}\label{Utilde-mn}
u^{I_1}_{\mu\nu}
\equiv
\frac{L}{2}
(\epsilon_\mu^{~\rho\sigma\lambda}\nabla_\nu s^{I_1}_{\rho\sigma\lambda}
+\epsilon_\nu^{~\rho\sigma\lambda}\nabla_\mu s^{I_1}_{\rho\sigma\lambda}).
\end{align}
We note that $\hat\psi^{I_1} \equiv \hat\psi^{I_1\mu}_{~~~\mu}$ and $\hat\phi^{I_1} \equiv \hat\phi^{I_1\mu}_{~~~\mu}$ are the gauge invariant scalar fields defined in \eqref{psiphi1}.

In order to define the 4-dimensional spin two physical modes, we write \eqref{phimn-eq} in terms of traceless tensor modes
\begin{align}
&
\Big(
\square
+\Lambda^{I_1}
+\frac{32}{L^2}
\Big)
\hat\psi^{I_1}_{(\mu\nu)}
+\frac{90}7
\Big(
\square
+\Lambda^{I_1}
+\frac{32}{L^2}
\Big)
\hat\phi^{I_1}_{(\mu\nu)}
-\frac{216}{7}\nabla_{(\mu}\nabla_{\nu)}\hat\phi^{I_1}
=
0,
\nn \\
&
\Big(
\square
+\Lambda^{I_1}
+\frac{44}{L^2}
\Big)
\hat\phi^{I_1}_{(\mu\nu)}
+\frac{14}{5L^2}\hat\psi^{I_1}_{(\mu\nu)}
=0,\label{hmn-eq7}
\end{align}
where $\hat\psi^{I_1}_{(\mu\nu)}=\hat\psi^{I_1}_{\mu\nu}-\frac14g_{\mu\nu}\hat\psi^{I_1}$ and $\hat\phi^{I_1}_{(\mu\nu)}=\hat\phi^{I_1}_{\mu\nu}-\frac14g_{\mu\nu} \hat\phi^{I_1}$.
Then we introduce the transverse and traceless spin two modes, which should satisfy the transverse condition $\nabla^{\mu}\nabla^{\nu}\hat h_{(\mu\nu)}^{I_1}=0$, and diagonalized linear equation,
\begin{align}\label{hmn-ans}
\check h_{(\mu\nu)}^{I_1}
=
\hat\phi^{I_1}_{(\mu\nu)}
+\frac7{30}\hat\psi^{I_1}_{(\mu\nu)}
+\frac{3L^2}{( I_1+2) (I_1+4)}\nabla_{(\mu}\nabla_{\nu)}\hat\phi^{I_1}
-\frac{7L^2}{30( I_1+2) (I_1+4)}\nabla_{(\mu}\nabla_{\nu)}\hat\psi^{I_1}.
\end{align}
Using \eqref{hmn-eq7}, we can write the linear equation for the diagonalized spin two KK modes as
\begin{align}\label{diagh}
(\square-M_{I_1}^2)
\check h_{(\mu\nu)}^{I_1}=0,
\end{align} 
where $M_{I_1}^2=\frac{I_1(I_1+6)-8}{L^2}$.

Similar to the spin zero case, the equation of motion for spin two modes contains higher derivative terms starting from quadratic order.
Therefore, one needs to introduce field redefinitions of the type \eqref{field-red1} in order to absorb those higher derivative terms.
Such field redefinition will result in the 4-dimensional spin two modes $ H_{(\mu\nu)}^{I_1}$.
However, at linear order, $H_{(\mu\nu)}^{I_1}=\check h^{I_1}_{(\mu\nu)}$ are the correct 4-dimensional spin two KK modes.

To summarize, the KK reduction of the bosonic sector of the 11-dimensional supergravity yields, three towers of scalar modes $ \Phi^{I_1}, \Psi^{I_1}, T^{I_{27}}$, two towers of pseudoscalar modes $ T_+^{I_{35}}$ and $T_-^{I_{35}}$, two towers of vector modes $V^{I_{7}}_{\mu}$ and $U^{I_{7}}_{\mu}$, one tower of pseudovector mode $T_\mu^{I_{21}}$, and one tower of spin-two mode $H^{I_1}_{(\mu\nu)}$.

\section{Exact KK Holography for LLM Geometries}\label{KKH-LLM}
In this section, we want to obtain exact results for the vevs of the gauge invariant operators in small mass expansion by using KK renormalization method~\cite{Skenderis:2006uy,Skenderis:2006di,Skenderis:2007yb} and compare with the field theory results of section \ref{VCPO-mABJM}.
At leading order in the small mass parameter, the linearized equations of motion discussed in the previous section are sufficient.
From the asymptotic expansions of the LLM geometries, one can read the solutions of some physical modes in 4-dimensional gravity, which are related to the vevs of gauge invariant operators.
For instance, the solutions of the KK mode, which are dual to the CPO with $\Delta = 1$, are completely determined from the leading asymptotic expansion of the LLM geometries.

\subsection{Asymptotic expansion of the LLM geometries}
As discussed in section \ref{mABJMLLM}, the LLM solutions are completely determined by two functions $Z(\tilde x,\tilde y)$ and $V(\tilde x,\tilde y)$ in \eqref{ZV}.
With some algebra these functions can be written in terms of the Legendre polynomials as follows
\begin{align}\label{ZVfunc}
&Z(\rho,\xi)=\frac1{2}\Big[\xi+\sum_{n=1}^{\infty}\big[(n+1)P_{n+1}(\xi)-2\xi nP_{n}(\xi)+(n-1)P_{n-1}(\xi)\big]C_n\big({\sqrt 2\mu_0 \rho}\big)^n\Big],\nn\\
&V(\rho,\xi)=\frac{2\rho}{L^3}\Big[1+\sum_{n=1}^{\infty}nP_{n}(\xi)C_n\big({\sqrt 2\mu_0 \rho}\big)^n\Big],
\end{align}
where $\rho=\frac{L^3}{4\tilde r}$, $\xi =\frac{\tilde x}{\tilde r}$ with $\tilde r = \sqrt{\tilde x^2+\tilde y^2}$, and $P_n(\xi)$ are the Legendre polynomials.
We have defined~\cite{Kim:2016dzw}
\begin{align}\label{Cn}
C_n = \sum_{i=1}^{2 N_B +1}(-1)^{i+1} \left(\frac{\tilde x_i}{2 \pi l_{\rm P}^3\mu_0 
\sqrt{A}}\right)^n,
\end{align}
where $A=kN-\sum_{j=1}^{N_B}\frac{t_{2j+1}(k-t_{2j+1})}{2}$ is the area in the Young diagram representation of the geometry and $t_i$'s are the discrete torsions assigned at the boundary $\tilde x_i$ between the black and white strips in the droplet picture~\cite{Cheon:2011gv}.
It can be shown that, the first two of the parameters $C_n$'s satisfy an identity:
\begin{align}\label{identity}
\beta_2\equiv C_2-C_1^2=2.
\end{align}

In terms of the $(\rho,\xi)$ coordinates, the metric of the LLM geometries in \eqref{LLMmet} is rewritten as
\begin{align}\label{LLM-met3}
ds^2
=
-{\bf G}_{tt}
\left(
-dt^2
+dw_1^2
+dw_2^2
\right)
+{\bf G}_{\rho\rho}
\left(
d\rho^2
+\frac{\rho^2}{1-\xi^2} d\xi^2
\right) 
+{\bf G}_{\theta \theta } ds_{S^3}^2 
+{\bf G}_{\tilde{\theta } \tilde{\theta }} ds_{\tilde{S}^3}^2 ,
\end{align}
where ${\bf G}_{\rho\rho} = \frac{L^6 {\bf G}_{xx}}{16 \rho^4}$.

To implement the method of holographic renormalization, we should rewrite the solution in terms of the Fefferman-Graham (FG) coordinate system~\cite{FGcoord}, where the holographic direction is well defined.
In a geometry which is asymptotically AdS$_{d+1}\times {\cal X}$, the metric in the FG coordinate system is given by
\begin{align}\label{FG-coord}
ds^2=\frac {L_{\rm AdS}^2}{z^2}\big(dz^2+{\bf g}_{ij}(z,y)dx^idx^j\big)+{\bf g}(z,y)ds^2_{\cal X},
\end{align}
where $z$ denotes the holographic direction and ${\bf g}(z,y)$ is some warp factor for the compact space.
In order to rewrite the LLM metric \eqref{LLM-met3} in the form \eqref{FG-coord} we use the coordinate transformations
\begin{align}\label{uxi_trans}
\rho = \rho(z,\tau), \qquad \xi = \xi(z,\tau),
\end{align}
which should satisfy two conditions
\begin{align}\label{two_condi}
&{\bf G}_{\rho\rho}\left(\frac{\partial \rho}{\partial z}\right)^2+\frac{\rho^2 {\bf G}_{\rho\rho}}{1-\xi ^2}\left(\frac{\partial \xi }{\partial z}\right)^2 =\frac{L^2}{4 z^2}, 
\nn \\
& {\bf G}_{\rho\rho}\left(\frac{\partial \rho}{\partial z}\right)\left(\frac{\partial \rho}{\partial \tau }\right)
+\frac{\rho^2 {\bf G}_{\rho\rho}}{1-\xi ^2}\left(\frac{\partial \xi }{\partial z}\right)\left(\frac{\partial \xi }{\partial \tau }\right)=0.
\end{align}
Then we obtain
\begin{align}\label{dsFG}
ds^2 =&\frac{ L^2}{4 z^2}\Big(dz^2 +\frac{4z^2}{L^2} {\bf g}_1(z,\tau )\left( -dt^2+dw_1^2+dw_2^2\right) \Big)\nn\\
&+ {\bf g}_2(z,\tau ) d\tau^2+ {\bf g}_3(z,\tau ) ds_{S^3}^2 + {\bf g}_4(z,\tau ) ds_{\tilde S^3}^2,
\end{align}
where
\begin{align}\label{FGCC}
&{\bf g}_1(z,\tau)={\bf G}_{tt}\big(\rho(z,\tau),\xi(z,\tau)\big),\nn\\
& {\bf g}_2(z,\tau)={\bf G}_{xx}\big(\rho(z,\tau),\xi(z,\tau)\big)\left[\Big(\frac{\partial \rho}{\partial \tau}\Big)^2+\frac{\rho^2}{1-\xi^2}\Big(\frac{\partial \xi}{\partial \tau}\Big)^2\right],\nn\\
&{\bf g}_3(z,\tau)={\bf G}_{\theta\theta}\big(\rho(z,\tau),\xi(z,\tau)\big),\qquad
{\bf g}_4(z,\tau)={\bf G}_{\tilde\theta\tilde\theta}\big(\rho(z,\tau),\xi(z,\tau)\big).
\end{align}
It is non-trivial to solve the two conditions in \eqref{two_condi} analytically.
However, recalling the asymptotic behavior of the LLM geometry, we note that the coordinate transformations satisfy the boundary conditions $\rho(z,\tau)|_{z\to 0} = z$ and $\xi (z,\tau)|_{z\to 0}= \tau$.
To solve the conditions \eqref{two_condi} in the asymptotic region, we use the following ansatze,
\begin{align}\label{uxi_trans2}
\rho(z,\tau) &= z\left(1+ a_1 z\mu_0 + a_2 z^2 \mu_0^2 + \cdots\right), 
\nn \\
\xi(z,\tau) &= \tau +b_1 z\mu_0 + b_2 z^2 \mu_0^2 + \cdots, 
\end{align}
where the $a_i$ and $b_i$ are determined from \eqref{two_condi},
\begin{align}
& a_1 =\frac{1}{3\sqrt2}(C_1^3-C_3)\tau,
\nn \\
&a_2 = \frac1{96}\Big(-8 C_2 C_1^4+12 C_3 C_1^3+3 \left(C_2^2-C_4\right) C_1^2-4 C_2 C_3 C_1+C_2^3-4 C_3^2+3 C_2 C_4\Big)\nn\\
&~~~~+\frac1{288}\Big(44 C_1^6-60 C_2 C_1^4-100 C_3 C_1^3+9 \left(11 C_2^2+5 C_4\right) C_1^2-12 C_2 C_3 C_1\nn\\
&~~~~~~~~~~~~~~~~-27 C_2^3+56 C_3^2-45 C_2 C_4\Big)\tau^2,\nn \\
&b_1 =-\frac{1}{3\sqrt2}(C_1^3-C_3)(1-\tau^2),
\nn \\
&b_2 =-\frac1{288}\Big(44 C_1^6-60 C_2 C_1^4-100 C_3 C_1^3+9 \left(11 C_2^2+5 C_4\right) C_1^2-12 C_2 C_3 C_1\nn\\
&~~~~~~~~~~~~~~~~-27 C_2^3+56 C_3^2-45 C_2 C_4\Big)\tau(1-\tau^2).
\end{align}

Using the above coordinate transformations, the asymptotic expansions of the warp factors ${\bf g}_{i}(z,\tau)$ are obtained and listed in appendix \ref{AstExp-app}.
The expansions depend on the $\tau$ coordinate on $S^7$ so that we can use the results in appendix \ref{Scalar-Harm} to replace the $\tau$ dependence in terms of the scalar harmonics on $S^7$. Then we read the values of $h_{ij}^{I_1}(i,j=t,w_1,w_2)$, for $I_1=0,2,4,\cdots$,
\begin{align}
&h^0_{ij}=\Big[-\frac{L^2\mu_0^2}{720}\big(17 C_1^6-51 C_2 C_1^4-28 C_3 C_1^3+72 C_2^2 C_1^2+42 C_2 C_3 C_1-45 C_2^3-7 C_3^2\big)+{\cal O}(\mu_0^4)\Big]\eta_{ij},\\
&h^2_{ij}=\Big[-\frac{L^2\mu_0}{3z}\big(2 C_1^3-3 C_2 C_1+C_3\big)+{\cal O}(\mu_0^3)\Big]\eta_{ij},\\
&h^{4}_{ij}=\Big[\frac{L^2\mu_0^2}{36\sqrt{10}}\big(28 C_1^6-84 C_2 C_1^4+28 C_3 C_1^3+9 \left(7 C_2^2-15 C_4\right) C_1^2+228 C_2 C_3 C_1\nn\\
&~~~~~~~~~~~~-135 C_2^3-128 C_3^2+135 C_2 C_4\big)+{\cal O}(\mu_0^4)\Big]\eta_{ij},
\end{align}
where $\eta_{ij} = {\rm diag}(-1,1,1)$.
Noting that $h_{zz}^{I_1} = 0$ in the FG coordinate, we obtain the values of $h^{I_1}\equiv g^{ij}h_{ij}^{I_1} = \frac{4z^2}{L^2} \eta^{ij} h_{ij}^{I_1}$,
\begin{align}
&h^0=-\frac{(\mu_0z)^2}{60}\big(17 C_1^6-51 C_2 C_1^4-28 C_3 C_1^3+72 C_2^2 C_1^2+42 C_2 C_3 C_1-45 C_2^3-7 C_3^2\big)+{\cal O}(\mu_0^4),\\
&h^2=-{4\mu_0z}\big(2 C_1^3-3 C_2 C_1+C_3\big)+{\cal O}(\mu_0^3),\\
&h^{4}=\frac{(\mu_0z)^2}{3\sqrt{10}}\big(28 C_1^6-84 C_2 C_1^4+28 C_3 C_1^3+9 \left(7 C_2^2-15 C_4\right) C_1^2+228 C_2 C_3 C_1\nn\\
&~~~~~~~~~~~~~~~~-135 C_2^3-128 C_3^2+135 C_2 C_4\big)+{\cal O}(\mu_0^4).
\end{align}
In appendix \ref{AstExp-app} we have also listed the expansions of the components of the 4-form field strength.
Using \eqref{F4-exp}, we read the values of the scalar fields $\tilde u^{I_1}$ defined in \eqref{Utilde} from $\epsilon^{\mu\nu\rho\sigma}f_{\mu\nu\rho\sigma}=4!\epsilon^{tw_1w_2z}f_{tw_1w_2z}$, where $f_{tw_1w_2z}$ is listed in appendix \ref{AstExp-app},
\begin{align}
&\tilde u^0=\frac{3(\mu_0z)^2}{10}\big(7 C_1^6-21 C_2 C_1^4+12 C_2^2 C_1^2+6 \left(2 C_1^2-3 C_2\right) C_3 C_1+5 C_2^3+3 C_3^2\big)+{\cal O}(\mu_0^4),\\
&\tilde u^2=-{48\mu_0z}\big(2 C_1^3-3 C_2 C_1+C_3\big)+{\cal O}(\mu_0^3),\\
&\tilde u^{4}=-\frac{2(\mu_0z)^2}{\sqrt{10}}\big(4 C_1^6-12 C_2 C_1^4+4 C_3 C_1^3+9 \left(C_2^2+15 C_4\right) C_1^2-276 C_2 C_3 C_1\nn\\
&~~~~~~~~~~~~~~~~+136 C_3^2+135 \left(C_2^3-C_2 C_4\right)\big)+{\cal O}(\mu_0^4).
\end{align}
In order to read the values of the scalars $\phi^{I_1}$, we take the trace of $h_{ab}$,
\begin{align}\label{haa}
h^a_{~a}=\phi^{I_1} Y^{I_1} = g^{ab}\big({\bf g}_{ab}-g_{ab}\big)=g^{ab}{\bf g}_{ab}-7.
\end{align}
Then the results are read from the asymptotic expansions of ${\bf g}_{3}(z,\tau)$ and ${\bf g}_{4}(z,\tau)$ in appendix \ref{AstExp-app},
\begin{align}
&\phi^0=\frac{(\mu_0z)^2}{15}\big(14 C_1^6-42 C_2 C_1^4+4 C_3 C_1^3+39 C_2^2 C_1^2-6 C_2 C_3 C_1-10 C_2^3+C_3^2\big)+{\cal O}(\mu_0^4),\\
&\phi^2={4\mu_0z}\big(2 C_1^3-3 C_2 C_1+C_3\big)+{\cal O}(\mu_0^3),\\
&\phi^{4}=\frac{(\mu_0z)^2}{3\sqrt{10}}\big(-124 C_1^6+372 C_2 C_1^4-124 C_3 C_1^3+9 \left(15 C_4-31 C_2^2\right) C_1^2-84 C_2 C_3 C_1\nn\\
&~~~~~~~~~~~~~~~~+104 C_3^2+135 \left(C_2^3-C_2 C_4\right)\big)+{\cal O}(\mu_0^4).
\end{align}
The values of the scalars $s^{I_1}$ are obtained from
\begin{align}\label{ddhexp}
\nabla^a\nabla^bh_{(ab)}=s^{I_1}\nabla^a\nabla^b\nabla_{(a}\nabla_{b)}Y^{I_1}=6s^{I_1}\Lambda^{I_1}\Big(\frac{\Lambda^{I_1}}7+\frac1{L^2}\Big)Y^{I_1},
\end{align}
by expanding,
\begin{align}\label{metexp2}
\nabla^a\nabla^bh_{(ab)}
=\nabla^a\nabla^b{\bf g}_{ab}-\frac17\square h^c_{~c}.
\end{align}
Then using appendix \ref{AstExp-app} we obtain
\begin{align}
&s^0=-\frac{L^2(\mu_0z)^2}{630}\big(14 C_1^6-42 C_2 C_1^4+4 C_3 C_1^3+39 C_2^2 C_1^2-6 C_2 C_3 C_1-10 C_2^3+C_3^2\big)+{\cal O}(\mu_0^4),\\
&s^2=\frac{L^2\mu_0z}{3}\big(2 C_1^3-3 C_2 C_1+C_3\big)+{\cal O}(\mu_0^3),\\
&s^{4}=\frac{L^2(\mu_0z)^2}{216\sqrt{10}}\big(-124 C_1^6+372 C_2 C_1^4-124 C_3 C_1^3+9 \left(15 C_4-31 C_2^2\right) C_1^2-84 C_2 C_3 C_1\nn\\
&~~~~~~~~~~~~~~~~+104 C_3^2+135 \left(C_2^3-C_2 C_4\right)\big)+{\cal O}(\mu_0^4).
\end{align}

To determine the graviton mode, in addition to $h_{\mu\nu}$ we also need the values of the tensor fields $\tilde u_{\mu\nu}$ defined in \eqref{Utilde-mn}.
We can rewrite the definition in \eqref{Utilde-mn}as
\begin{align}\label{Utilde-mn2}
\tilde u^{I_1}_{\mu\nu}\Lambda^{I_1}Y^{I_1}=\frac L2\big(\epsilon_\mu^{~~\rho\sigma\lambda}\nabla_\nu s^{I_1}_{\rho\sigma\lambda}\Lambda^{I_1}Y^{I_1}+\epsilon_\nu^{~~\rho\sigma\lambda}\nabla_\mu s^{I_1}_{\rho\sigma\lambda}\Lambda^{I_1}Y^{I_1}\big).
\end{align}
On the other hand, from \eqref{F4-exp}, we see that $\nabla^a f_{\mu\nu\rho a}=-s^{I_1}_{\mu\nu\rho}\Lambda^{I_1} Y^{I_1}$.
Using this into \eqref{Utilde-mn2} and noting that, for the LLM geometry, the only non zero $f_{\mu\nu\rho a}$ is $f_{tw_1w_2\tau}$, we obtain the following results
\begin{align}\label{Dfmnra-exp4}
\tilde u^{I_1}_{ij}\Lambda^{I_1} Y^{I_1}
=&3!\frac{4z}{L}\Big(\frac{4(1-\tau^2)}{L^2}\partial_\tau F_{tw_1w_2\tau}-\frac{16\tau}{L^2}F_{tw_1w_2\tau}\Big)\eta_{ij},
\\
\tilde u^{I_1}_{zz}\Lambda^{I_1} Y^{I_1}
=&-3!\frac{4z^2}{L}\Big[\frac{4(1-\tau^2)}{L^2}\big(\partial_\tau\partial_z F_{tw_1w_2\tau}+3\frac1z\partial_\tau F_{tw_1w_2\tau}\big)-\frac{16\tau}{L^2}\big(\partial_z F_{tw_1w_2\tau}+3\frac1zF_{tw_1w_2\tau}\big)\Big].
\nn
\end{align}
Using again the results of appendix \ref{AstExp-app}, we read the values of $\tilde u^{I_1}_{\mu\nu}$, for $I_1=2,4,\cdots$.
However, since $\Lambda^0=0$, we can not read $\tilde u^{0}_{\mu\nu}$.
\begin{align}
&\tilde u^2_{ij}=\Big(-\frac{6L^2\mu_0}{z}\big(2C_1^3-3C_1C_2+C_3\big)+{\cal O}(\mu_0^3)\Big)\eta_{ij},\\
&\tilde u^{4}_{ij}=\Big(-\frac{L^2\mu_0^2}{2\sqrt{10}}\big(4 C_1^6-12 C_2 C_1^4+4 C_3 C_1^3+9 \left(C_2^2+15 C_4\right) C_1^2-276 C_2 C_3 C_1\\
&~~~~~~~~~~~~~~~~+135 C_2^3+136 C_3^2-135 C_2 C_4\big)+{\cal O}(\mu_0^4)\Big)\eta_{ij},\nn\\
&\tilde u^2_{zz}=\frac{6L^2\mu_0}{z}\big(2C_1^3-3C_1C_2+C_3\big)+{\cal O}(\mu_0^3),\\
&\tilde u^{4}_{zz}=\frac{L^2\mu_0^2}{\sqrt{10}}\big(4 C_1^6-12 C_2 C_1^4+4 C_3 C_1^3+9 \left(C_2^2+15 C_4\right) C_1^2-276 C_2 C_3 C_1\\
&~~~~~~~~~~~~~~~~+135 C_2^3+136 C_3^2-135 C_2 C_4\big)+{\cal O}(\mu_0^4).\nn
\end{align}
Finally, the vector fields $s_\mu^{I_1}$, which are also needed to write the graviton mode, are zero, because for the LLM geometry $h_{\mu a}$ is zero.

\subsection{Asymptotic expansions for the physical modes in 4 dimensions}
The asymptotic expansions for the scalar and tensor fluctuations in the previous subsection are at least linear in the mass parameter $\mu_0$.
If we truncate our results at $\mu_0$ order, the only non-vanishing fields are the KK modes with $I_1=2$.
Keeping this in mind, we list the linear order asymptotic expansions of some of the scalar and tensor physical modes discussed in subsection \ref{phys-modes}.

Plugging the expansions of the previous subsection into \eqref{scme1}, we obtain the asymptotic expansions for $\Phi^{I_1}$ and $\Psi^{I_1}$,
\begin{align}\label{field-red6}
&\Phi^{0} = {\cal O}(\mu_0^2), \qquad\qquad \Phi^{2} = {\cal O}(\mu_0^3), \qquad\qquad\Phi^{4} = {\cal O}(\mu_0^2),\nn
\\
& \Psi^{0} = {\cal O}(\mu_0^2), \qquad \Psi^{2} = -24\beta_3\mu_0 z+{\cal O}(\mu_0^3), \qquad\Psi^{4} = {\cal O}(\mu_0^2),
\end{align}
where
\begin{align}\label{beta1}
\beta_3 \equiv 2C_1^3-3C_1C_2+C_3. 
\end{align}
The solutions in \eqref{field-red6} satisfy the linearized equation of motion \eqref{diagpp}.

The asymptotic expansions of the remaining three spin zero modes, $ T^{I_{27}}, ~ T_+^{I_{35}},~ T_-^{I_{35}}$, in \eqref{Extra-Sca} cannot be determined without having the explicit form of vector and tensor spherical harmonics.
However, those spin zero modes are not needed for our purpose here for the following reason.
In AdS/CFT correspondence~\cite{Gubser:1998bc,Witten:1998qj}, the mass $m^2$ of a scalar field on the gravity side is related to the conformal dimension $\Delta$ of the dual gauge invariant operator by
\begin{align}
m^2R^2_{AdS_{d+1}}=\Delta(\Delta-d).
\end{align}
In our case ($R^2_{AdS_{4}}=\frac{L^2}4$), where $L$ is the $S^7$ radius, we have
\begin{align}\label{mL}
m^2L^2=2\Delta(2\Delta-6).
\end{align}
Then the conformal dimensions of the gauge invariant operators, which are dual to the spin zero fields are as follows:
\begin{itemize}
\item For the scalar field $\Psi^{I_1}$ we have
\begin{align}
{I_1(I_1-6)}=2\Delta(2\Delta-6)\Rightarrow \Delta=\frac{I_1}2, \quad \{I_1=2,4,6,\cdots\}.
\end{align}
\item For the scalar field $\Phi^{I_1}$ we have
\begin{align}
{(I_1+12)(I_1+6)}=2\Delta(2\Delta-6)\Rightarrow \Delta=\frac{I_1+12}2, \quad \{I_1=0,2,4,\cdots\}.
\end{align} 
\item For the scalar field $ T^{I_{27}}$ we have
\begin{align}
{(I_{27}+6)I_{27}}=2\Delta(2\Delta-6)\Rightarrow \Delta=\frac{I_{27}+6}2, \quad \{I_{27}=2,4,6\cdots\}.
\end{align} 
\item For the pseudoscalar field $ T^{I_{35}}_{-}$ we have
\begin{align}
{(I_{35}+3)(I_{35}-3)}=2\Delta(2\Delta-6)\Rightarrow \Delta=\frac{I_{35}+3}2, \quad \{I_{35}=1,3,5,\cdots\}.
\end{align} 
\item For the pseudoscalar field $ T^{I_{35}}_{+}$ we have
\begin{align}
{(I_{35}+9)(I_{35}+3)}=2\Delta(2\Delta-6)\Rightarrow \Delta=\frac{I_{35}+9}2, \quad \{I_{35}=1,3,5,\cdots\}.
\end{align} 
\end{itemize}
In the above list, $ T_+^{I_{35}}$ and $T_-^{I_{35}}$ are pseudoscalars and cannot be the candidates for the dual scalar fields of CPOs.
In addition, if we are interested in CPOs of lower conformal dimensions $\Delta=1,2,\cdots$, then $\Phi^{I_1}$ and $ T^{I_{27}}$ are also not the candidates.
The only scalar fields which can be dual to CPOs with lower conformal dimensions are $\Psi^{I_1}$, with $I_1=2,4,\cdots$.
For a similar reason we also skip the asymptotic expansion of the vector fields in subsection \ref{Vect-Eq}.

The asymptotic expansions of the spin two KK modes are needed to obtain the vev of the energy-momentum tensor in mABJM theory.
Following the same procedure as the spin zero modes, from \eqref{hmn-ans} we obtain the asymptotic expansions of the spin two modes
\begin{align}\label{field-red-hmn}
H^{2}_{(\mu\nu)} = {\cal O}(\mu_0^3), \qquad H^{4}_{(\mu\nu)} = {\cal O}(\mu_0^4).
\end{align}
At $\mu_0$ order, we obtain a result which is consistent with the fact that the mABJM theory is a supersymmetric theory and the vev of the energy-momentum tensor is vanishing.
Furthermore, if we go beyond the linear order, the spin two KK modes are modified by field redefinition \eqref{field-red1} and the vev of energy-momentum tensor should still be vanishing.

\subsection{Comparison with field theory results}
According to the holographic renormalization procedure~\cite{Henningson:1998gx}, the vev of a CPO with conformal dimension $\Delta$ is determined by the coefficient $\phi_{(\Delta)}$ of $z^{\Delta}$ in the asymptotic expansion of of a dual gauge invariant scalar fields $\Phi$ on the gravity side, i.e.
\begin{align}\label{VEV-CPO2}
\langle{\cal O}^{(\Delta)}\rangle_m =\frac{N^2}{\sqrt{\lambda}}{\mathbb N} \phi_{(\Delta)}
\end{align}
where ${\mathbb N}$ is a numerical number depending on the normalization of the dual scalar field in 11-dimensional supergravity,  and $\lambda$ is defined as $\lambda = N/k$ in the ABJM theory.  In the case $k=1$, the overall normalization in \eqref{VEV-CPO2} is reduced to $N^{\frac32}$. 
The $N^2/\sqrt{\lambda}$-dependence in the right-hand side 
of \eqref{VEV-CPO2} is a peculiar behavior of the normalization 
factor in holographic dual relation 
for the M2-brane theory~\cite{Klebanov:1996un,Aharony:2008ug,Drukker:2010nc}. 

In the pervious subsection, we have shown that only the scalar modes $\Psi^{I_1}$ have non-trivial asymptotic expansions at linear order in $\mu_0$.
At quadratic order or higher, more of the scalar modes as well as the spin two modes have non-trivial asymptotic expansions.
On the other hand, holographic renormalization states that the dual operators of these extra modes should have vanishing vevs.
In order to reconcile these differences we need the field redefinition~\cite{Skenderis:2006uy} of the type \eqref{field-red1} to obtain the correct 4-dimensional fields.
The field redefinition makes the asymptotic expansions of all the fields trivial except for $\Psi^{I_1}$.
We have obtained the asymptotic expansion of the scalar field $\Psi^{2}$ in \eqref{field-red6} while the vev of the dual operator was calculated in the subsection \ref{vev-CPO}.
Using the gauge/gravity duality relation in \eqref{VEV-CPO2} and setting $k=1$, these two results are related as
\begin{align}\label{VEV-CPO2-1}
\langle {\cal O}^{(1)}\rangle_m = \frac{N^2}{\sqrt{\lambda}}{\mathbb N}\, \psi^{(1)} 
= -  \frac{24 N^2}{\sqrt{\lambda}}\, {\mathbb N} \, \beta_3 \mu_0. 
\end{align}
In order to fix the normalization factor ${\mathbb N}$, we use the identity
\begin{align}\label{Identity}
\sum_{n=0}^{\infty} \Big[n(n+1)(l_n - l_n') \Big]
= \frac{N^{3/2}}3\beta_3,
\end{align}
which is proved in appendix \ref{Cpt1}. Then using \eqref{vevD1} we rewrite the duality relation \eqref{VEV-CPO2-1} as
\begin{align}\label{vevD2}
\frac{\mu }{4\sqrt2\pi}\sum_{n=0}^{\infty}\Big[n(n+1)(N_n-N_n')\Big]= -72{\mathbb N}\mu_0 \sum_{n=0}^{\infty}\Big[n(n+1)(l_n - l_n')\Big] .
\end{align}
Now recalling that $\mu=4\mu_0$ and using the one-to-one correspondence \eqref{NNn} between the occupation numbers of the vacua in mABJM theory and the discrete torsions in the LLM geometries, the normalization factor is fixed as ${\mathbb N} = -\frac{ \sqrt{2}}{144 \pi}$.
Therefore, the vev of the CPO with conformal dimension $\Delta=1$ is given by
\begin{align}\label{vevCPO3}
\langle {\cal O}^{(1)}\rangle_m =\frac{N^{\frac32}\mu_0}{3\sqrt{2}\,\pi}\, \beta_3.
\end{align}
In Young-diagram picture, $\beta_3$ depends on the shape of a Young-diagram but not on the size of the diagram, which means it is independent of the number of M2-branes.
The result \eqref{vevCPO3} is obtained for all possible 
supersymmetric vacua with large $N$ in the mABJM theory. The $N^{\frac32}$-dependence in the right-hand side 
of \eqref{vevCPO3} exactly matches the $N$ dependence of the 11-dimensional Newton's constant~\cite{Klebanov:1996un}, which is fixed by the Bekenstein-Hawking entropy formula. As it is expected by the holographic relation, this $N^{\frac32}$ behavior also agrees with the total number of degrees of freedom of the M2-brane theory in large $N$ limit~\cite{Drukker:2010nc}. 

The extension of the above results to $k\ne1$ case is not straightforward.
We postpone such extension in most general setup to future work.
Instead, here we extend the results for the LLM geometries with rectangular shaped Young-diagram representations.
In this case the droplet representation has only one finite size black strip i.e. $N_B=1$.
The exact dual relation is then given by
\begin{align}\label{vevCPOnek}
\langle {\cal O}^{(1)}\rangle_m =\frac{N \sqrt{k\tilde N}\mu_0}{3\sqrt{2}\,\pi}\, \beta_3
= \frac{N \sqrt{N\tilde N}\mu_0}{3\sqrt{2}\,\pi\,\sqrt{ \lambda}}\, \beta_3,
\end{align}
where $\tilde N = A/k$ with $A$ defined in \eqref{Cn} and we have introduced the 't Hooft coupling constant $\lambda= N/k$ in the ABJM theory.
In the large $N$ limit, $\tilde N$ is reduced to $N$ in \eqref{vevCPOnek}.

\section{Conclusion}\label{CONC}
In this paper, we have calculated the vevs of the CPO with conformal dimension $\Delta=1$ from all discrete supersymmetric vacua of mABJM theory as well as from the dual LLM solutions of 11-dimensional supergravity and found an exact holographic relation between the two results with $k=1$ in the large $N$ limit. 
Due to computational difficulties, we treated the case $k\ne1$ only for the LLM geometry with rectangular Young diagram.
On the field theory side, the vevs of CPOs are protected from quantum corrections due to the high supersymmetry of the mABJM theory and as a result they are completely determined by the supersymmetric vacuum solutions in the large $N$ limit.
The CPOs are given by single traces of products of the complex scalar fields in the mABJM and their vevs are obtained by evaluating those traces at the discrete supersymmetric vacua, which are represented by the GRVV matrices.
On the gravity side, the gauge invariant 4-dimensional scalar fields which are dual to the CPOs were obtained from the KK reduction of 11-dimensional supergravity.
We showed that the gauge invariant fields obtained from the KK reduction of the 11-dimensional LLM solutions satisfy the 4-dimensional equations of motion.
The equations of motion are satisfied order by order when we expand in powers of the mass parameter $\mu_0$. This expansion coincides with the asymptotic expansion in the holographic coordinate $z$ from which we read the vevs of the dual gauge invariant operators.
For the CPO with $\Delta=1$, the vevs are given by the first order terms in the $\mu_0$ expansion of the dual gauge invariant scalar fields.
We have also carried out this procedure for the energy-momentum tensor in mABJM theory and its vev is vanishing.
This result is expected because the theory is supersymmetric.

It seems that the discrete Higgs vacua of the mABJM theory as well as the corresponding LLM geometries are parametrized by the vevs of CPOs.
In other words, knowing the vevs of enough number of CPOs, one can fully determine the shape of the droplet picture of the LLM geometry and hence the discrete Higgs vacua of the mABJM theory.
For instance, the shape of the droplet with $N_B=1$ is fixed by the value of the vevs of the CPO with $\Delta=1$ as follows.
As it was discussed in subsection \ref{LLMgeom} the shape of the droplet is fixed by the values of $\tilde x_i$ with $i=1,\cdots,2N_B+1$.
For $N_B=1$, the values of $\tilde x_1,\tilde x_2$, and $\tilde x_3$ are determined by $C_1,C_2$, and $C_3$.
For our coordinate choice $C_1=0$, $C_2=2$ from the identity \eqref{identity} while $C_3$ is determined by the vevs of CPO with $\Delta=1$.
Therefore, the supersymmetric vacua corresponding to the droplet with $N_B=1$ are parametrized by the vevs of CPO with $\Delta=1$.
This is a meaningful result because, if we can calculate the vevs of enough number of CPOs, it is possible to project out the supersymmetric vacua from the full set of classical Higgs vacua in the mABJM theory at a given $N_B$.
Those supersymmetric vacua are in one-to-one correspondence with the half-BPS LLM solutions.
Our quantitative results for the gauge/gravity correspondence  contains partition of $N$ different cases. However, we need to accumulate more analytic evidences for CPOs with $\Delta$ ($\ge 2$) and $k$ ($\ge 1$) to completely determine the supersymmetric vacua at arbitrary $N_B$. We leave these issues for future study~\cite{JKKT}.

Recently, it is reported that for the mABJM theory on $S^3$, there is no gravity dual for the mass parameter larger than a critical mass value~\cite{Nosaka:2016vqf}.
See also \cite{Anderson:2014hxa,Anderson:2015ioa,Nosaka:2015bhf}.
Though the setup is different from ours, where the mABJM theory is defined on $R^{2,1}$, it is intriguing to investigate the gravity dual for our case in the large mass region and compare the results with those of mABJM theory on $S^3$.


\section*{Acknowledgements}
This work was supported by the National Research Foundation of Korea(NRF) grant with grant number NRF-2014R1A1A2057066 and NRF-2016R1D1A1B03931090 (Y.K.), and NRF-2014R1A1A2059761 (O.K.).

\appendix 
\section{Spherical Harmonics on $S^7$}\label{SHS7}
The spherical harmonics on $S^7$ are defined as follows
\begin{align}\label{SHdef}
&\nabla^2Y^{I_1}=\Lambda^{I_1}Y^{I_1}=-\frac{1}{L^2}I_1(I_1+6)Y^{I_1},~~{\rm Scalar},\nn\\
&\nabla^2Y_a^{I_7}=\Lambda^{I_7}Y_a^{I_7}=-\frac{1}{L^2}(I_7^2+6I_7-1)Y_a^{I_7},~~{\rm Vector},\nn\\
&\nabla^2Y_{(ab)}^{I_{27}}=\Lambda^{I_{27}}Y_{(ab)}^{I_{27}}=-\frac{1}{L^2}(I_{27}^2+6I_{27}-2)Y_{(ab)}^{I_{27}},~~{\rm Symmetric ~2-Tensor},\nn\\
&\nabla^2Y_{[ab]}^{I_{21}}=\Lambda^{I_{21}}Y_{[ab]}^{I_{21}}=-\frac{1}{L^2}(I_{21}^2+6I_{21}-2)Y_{[ab]}^{I_{21}},~~{\rm antiSymmetric ~2-Tensor},\nn\\
&\nabla^2Y_{[abc]}^{I_{35}}=\Lambda^{I_{35}}Y_{[abc]}^{I_{35}}=-\frac{1}{L^2}(I_{35}^2+6I_{35}-3)Y_{[abc]}^{I_{35}},~~{\rm antiSymmetric ~3-Tensor},\nn\\
&\nabla^a Y_a^{I_7}=\nabla^a Y_{(ab)}^{I_{27}}=\nabla^a Y_{[ab]}^{I_{21}}=\nabla^a Y_{[abc]}^{I_{35}}=0,\nn\\
&g^{ab} Y_{(ab)}^{I_{27}}=g^{ab} Y_{[ab]}^{I_{21}}=g^{ab} Y_{[abc]}^{I_{35}}=0.
\end{align}
Here, when we write $Y^{I_n}$, the subscript $n$ denotes the number of components of the spherical harmonics.

\subsection{Scalar Spherical Harmonics}\label{Spher-Harm}
The scalar spherical harmonics on $S^7$ are the restriction of
\begin{align}\label{YI1}
Y^{I_1}=\frac1{L^{I_1}}C^{I_1}_{i_1\cdots i_{I_1}}x^{i_1}\cdots x^{i_{I_1}}
\end{align}
to a seven-sphere, where $x^i$ with $(i=1,\cdots,8)$ are the Cartesian coordinates of ${\rm I\!R}^8$ and the coefficients $C^{I_1}_{i_1\cdots i_{I_1}}$ are totally symmetric and traceless.
In order to evaluate integrals which are quadratic or cubic in the scalar harmonics, we use the following general formula
\begin{align}
\frac1{\omega_7}\int_{S^7}x^{i_1}\cdots x^{i_{2m}}=\frac{3L^{2m}}{2^{m-1}(m+3)!}\big({\rm all~possible~pairing}\big),
\end{align}
where {\it all possible pairing} means, $\delta^{i_1i_2}$ for $m=1$ , $(\delta^{i_1i_2}\delta^{i_3i_4}+\delta^{i_1i_3}\delta^{i_2i_4}+\delta^{i_1i_4}\delta^{i_2i_3})$, for $m=2$ and a similar contraction for higher $m$.

Now consider
\begin{align}
\frac1{\omega_7}\int_{S^7}Y^{I_1}Y^{J_1}=\frac1{\omega_7L^{I_1+J_1}}\int_{S^7}C^{I_1}_{i_1\cdots i_{I_1}}C^{J_1}_{j_1\cdots j_{J_1}}x^{i_1}\cdots x^{i_{I_1}}x^{j_1}\cdots x^{j_{J_1}},
\end{align}
where $\omega_7=\frac{\pi^4}3$ is the surface area of $S^7$.
Recalling that the $C^{I_1}$ are traceless and also that the integral vanishes if any of the $x^i$ left un-paired, we note that we get a non-zero value only when $I_1=J_1$
\begin{align}
\frac1{\omega_7}\int_{S^7}Y^{I_1}Y^{J_1}=&\frac1{\omega_7L^{2I_1}}\int_{S^7}C^{I_1}_{i_1\cdots i_{I_1}}C^{J_1}_{j_1,\cdots j_{J_1}}x^{i_1}\cdots x^{i_{I_1}}x^{j_1}\cdots x^{j_{J_1}}\nn\\
&=\frac{3}{2^{I_1-1}(I_1+3)!}C^{I_1}_{i_1,\cdots i_{I_1}}C^{J_1}_{j_1\cdots j_{J_1}}\big({\rm all~possible~pairing}\big).
\end{align}
Because of the tracelessness of $C^{I_1}$, when we sum over the sets of indices $i$ and $j$, we get contributions only from the terms in which the $i$ indices are paired with the $j$ indices.
Those types of pairing result in a complete contraction of the $C^{I_1}$ and $C^{J_{1}}$ indices and the total number of such terms is $I_1!$.
Therefore we get
\begin{align}
\frac1{\omega_7}\int_{S^7}Y^{I_{1}}Y^{J_{1}}=\frac{3I_{1}!}{2^{I_{1}-1}(I_{1}+3)!}\langle C^{I_{1}}C^{J_{1}}\rangle,
\end{align}
where $\langle C^{I_{1}}C^{J_{1}}\rangle=C^{I_{1}}_{i_1\cdots i_{I_{1}}}C^{J_{1}}_{i_1\cdots i_{I_{1}}}$.
Actually, we normalize the scalar harmonics such that $\langle C^{I_{1}}C^{J_{1}}\rangle=\delta^{I_{1}J_{1}}$.
Expressions involving derivatives can be evaluated by using integration by parts.

\subsection{Vector Spherical Harmonics}
Consider a vector field in ${\rm I\!R}^8$
\begin{align}
Y^{I_{7}}_p=\frac1{L^{I_{7}}}V^{I_{7}}_{p,i_1\cdots i_{I_{7}}}x^{i_1}\cdots x^{i_{I_{7}}}, \quad p=1,\cdots 8
\end{align}
with $V^{I_{7}}_p$ traceless and totally symmetric in the $i_1,\cdots,i_{I_{7}}$ indices. It also satisfy $x^pY^{I_{7}}_p=0$.\footnote{We will use the index notation where $p,q,\cdots$ are the ${\rm I\!R}^8$ indices and $a,b,\cdots$ are the $S^7$ indices.}
The vector spherical harmonics $Y^{I_{7}}_a$ on $S^7$ is defined as the component of such vector field tangent to the $S^7$ with the $x^i$ are restricted to the sphere.
This can be written as
\begin{align}\label{YaI7}
Y^{I_{7}}_a=\hat e^p_aY^{I_{7}}_p=\frac1{L^{I_{7}}}\hat e^p_aV^{I_{7}}_{p,i_1\cdots i_{I_{7}}}x^{i_1}\cdots x^{i_{I_{7}}}, \quad a=1,\cdots,7,
\end{align}
where $\hat e^p$ is the unit vector along the $x^p$ Cartesian coordinates of ${\rm I\!R}^8$ and $\hat e^p_a$ is its projection onto the $a^{th}$ unit vector tangent to $S^7$.
Those projections are actually given by
\begin{align}
\hat e^p_a=\frac{\partial x^p}{\partial\theta^a},
\end{align}
where $\theta^a$ are the coordinates of $S^7$.
In order to evaluate the integrals involving those vector harmonics we use the following procedure.
Lets consider
\begin{align}
\int_{S^7}V_ag^{ab}V_b
\end{align}
for any two vectors $V_a, V_b$ tangent to $S^7$ and $g_{ab}$ is the metric on $S^7$.
Using the definitions in \eqref{YaI7} we can write
\begin{align}
\int_{S^7}V_ag^{ab}V_b=\int_{S^7}V_p\hat e^p_ag^{ab}\hat e^q_bV_q.
\end{align}
Now we can make the following replacement
\begin{align}\label{VV}
\hat e^p_ag^{ab}\hat e^q_b=\gamma^{pq}=\delta^{pq}-n^pn^q,
\end{align}
where $n^p=\frac{x^p}{L}$ are the eight components of the unit normal vector to $S^7$.
This relation can easily be verified by using the standard polar coordinates in ${\rm I\!R}^8$.
Therefore we can write
\begin{align}
\int_{S^7}V_ag^{ab}V_b=\int_{S^7}V_p(\delta^{pq}-n^pn^q)V_q.
\end{align}
Now we can evaluate the integrals of the vector harmonics by using the above results and following the same procedure as those of the scalar harmonics.
Lets see few examples
\begin{align}
\frac1{\omega_7}\int_{S^7}g^{ab}Y_a^{I_{7}}Y_b^{J_{7}}=&\frac1{\omega_7L^{I_{7}+J_{7}}}\int_{S^7}(\delta^{pq}-n^pn^q)V^{I_{7}}_{p,i_1\cdots i_{I_{7}}} V^{J_{7}}_{q,j_1\cdots j_{I_{7}}}x^{i_1}\cdots x^{i_{I_{7}}}x^{j_1}\cdots x^{j_{I_{7}}}.
\end{align}
Recalling that $x^pY^{I_{7}}_p=0$, the second piece is zero.
In general we can drop the second term in \eqref{VV} if at least one of the two vectors involved is a vector spherical harmonics.
Therefore we have
\begin{align}
\frac1{\omega_7}\int_{S^7}g^{ab}Y_a^{I_{7}}Y_b^{J_{7}}=&\frac1{\omega_7L^{2I_{7}}}\int_{S^7}V^{I_{7}}_{p,i_1\cdots i_{I_{7}}} V^{J_{7}}_{p,j_1\cdots j_{I_{7}}}x^{i_1}\cdots x^{i_{I_{7}}}x^{j_1}\cdots x^{j_{I_{7}}}=\frac{3I_{7}!}{2^{I_{7}-1}(I_{7}+3)!}\langle V^{I_{7}}V^{J_{7}}\rangle.
\end{align}
Tensor spherical harmonics are also treated using procedures similar to that of vector spherical harmonics.

\subsection{Scalar spherical harmonics on $S^{7}$ with $SO(4)\times SO(4)$ symmetry}\label{Scalar-Harm}
In dealing with scalar equations of motion, we mainly need the scalar spherical harmonics which satisfy the following harmonic equation
\begin{align}
\left(\nabla^2+\frac{I_{1}(I_{1}+6)}{L^2}\right)Y^{I_{1}}=\left(\frac1{\sqrt{g}}\partial_a\big(\sqrt{g}g^{ab}\partial_b\big)+\frac{I_{1}(I_{1}+6)}{L^2}\right)Y^{I_{1}}=0,
\end{align}
where $g_{ab}$ is the metric on $S^7$.
We introduce the following coordinate on $S^7$
\begin{align}
ds^{2}_{S^{7}}=L^2\left(d\theta^2+{\rm cos}^2\theta ds^{2}_{S^{3}}+{\rm sin}^2\theta ds^{2}_{\tilde S^{3}}\right).
\end{align}
In obtaing the AdS${}_4\times S^7$ as asymptotic limit of LLM geometry we saw that the $\theta$ coordinate of $S^7$ is related to the $\alpha$ coordinate of LLM as $\theta=\alpha/2$.
Actually, it is better use the coordinate $\tau=\cos\alpha$.
Then we have
\begin{align}
ds^{2}_{S^{7}}=L^2\left(\frac1{4(1-\tau^2)}d\tau^2+\frac{1+\tau}2 ds^{2}_{S^{3}}+\frac{1-\tau}2 ds^{2}_{\tilde S^{3}}\right).
\end{align}
Imposing the $SO(4)\times SO(4)$ symmetry, the spherical harmonics depend only on the $\tau$ coordinate, which implies
\begin{align*}
\left(\frac1{\sqrt{g}}\partial_a\big(\sqrt{g}g^{ab}\partial_b\big)+\frac{I_{1}(I_{1}+6)}{L^2}\right)Y^{I_{1}}=0
\Longrightarrow\left[(1-\tau^2)\partial_\tau^2-4\tau\partial_\tau+\frac{I_{1}(I_{1}+6)}4\right]Y^{I_{1}}(\tau)=0.
\end{align*}
This is the hypergeometric equation with the following two independent solutions
\begin{align}
Y^{I_{1}}(\tau)=N^{I_{1}}~_2F_1\left(-\frac {I_{1}}4,\frac{I_{1}+6}4,\frac12;\tau^2\right),\quad Y^{I_{1}}(\tau)=N^{I_{1}}~_2F_1\left(-\frac {I_{1}-2}4,\frac{I_{1}+8}4,\frac32;\tau^2\right).
\end{align}
For $I_{1}=4i$, $(i=0,1,2,\cdots)$, the first solution is a polynomial and the first few terms are
\begin{align}
&Y^0=1,\quad Y^4=\frac1{8\sqrt{10}}(1-5\tau^2),\quad \cdots.
\end{align}
For later convenience lets invert these relation and write the following
\begin{align}\label{inverse-SH}
& \tau^2=\frac{Y^0-8\sqrt{10}~Y^4}5,\quad\cdots.
\end{align}
On the other hand, for $I_{1}=4i+2$, $(i=0,1,2,\cdots)$, the second solution is a polynomial and the first few terms are
\begin{align}\label{Y2}
&Y^2=\frac1{2\sqrt2}\tau,\quad\cdots,
\end{align}
which gives
\begin{align}
&\tau=2\sqrt{2}~Y^2,\quad \cdots.
\end{align}

\subsection{$C^{I_{1}=2}$ and $C^{(\Delta=1)}$}\label{CabD1}
In subsection \ref{CPOs-def}, we have stated the fact that the coefficients $C^{I_{1}}_{i_1\cdots i_{I_{1}}}$, which defines the scalar spherical harmonics in \eqref{YI1} are related to the coefficients $C^{(\Delta)A_1\cdots A_{n}}_{B_1\cdots B_{n}}$ of the CPOs in \eqref{CPO}.
In this appendix, we determine these coefficients in the particular case of CPO with conformal dimension $\Delta=1$.
First lets determine $C^{2}_{ij}$ from \eqref{YI1} and \eqref{Y2}.
Since the solution in \eqref{Y2} is obtained by imposing the $SO(4)\times SO(4)$ symmetry, we rewrite the scalar harmonics in \eqref{YI1} in a form that manifests this symmetry
\begin{align}\label{Y22}
Y^{2}=\frac1{L^2}\Big(\sum_{i,j=1}^{4}C^2_{ij}x^{i}x^{j}+\sum_{i,j=5}^{8} C^2_{ij}x^{i}x^{j}\Big),
\end{align}
with the ${\mathbb R}^8$ coordinates restricted to $S^7$ are written as follows
\begin{align}
&x^1=L\Big(\frac{1+\tau}2\Big)^{\frac12}\cos\big(\frac\theta2\big)\cos\big(\frac{\phi+\psi}2\big),\quad x^2=L\Big(\frac{1+\tau}2\Big)^{\frac12}\cos\big(\frac\theta2\big)\sin\big(\frac{\phi+\psi}2\big),\nn\\
&x^3=-L\Big(\frac{1+\tau}2\Big)^{\frac12}\sin\big(\frac\theta2\big)\sin\big(\frac{\phi-\psi}2\big),\quad
x^4=L\Big(\frac{1+\tau}2\Big)^{\frac12}\sin\big(\frac\theta2\big)\cos\big(\frac{\phi-\psi}2\big),\nn\\
&x^5=L\Big(\frac{1-\tau}2\Big)^{\frac12}\cos\big(\frac{\tilde\theta}2\big)\cos\big(\frac{\tilde\phi+\tilde\psi}2\big),\quad x^6=L\Big(\frac{1-\tau}2\Big)^{\frac12}\cos\big(\frac{\tilde\theta}2\big)\sin\big(\frac{\tilde\phi+\tilde\psi}2\big),\nn\\
&x^7=-L\Big(\frac{1-\tau}2\Big)^{\frac12}\sin\big(\frac{\tilde\theta}2\big)\sin\big(\frac{\tilde\phi-\tilde\psi}2\big),\quad
x^8=L\Big(\frac{1-\tau}2\Big)^{\frac12}\sin\big(\frac{\tilde\theta}2\big)\cos\big(\frac{\tilde\phi-\tilde\psi}2\big).
\end{align}
From \eqref{Y2} we notice that $Y^2$ depends only on the $\tau$ coordinate of $S^7$.
Therefore, the dependence on the remaining $S^7$ coordinates should disappear in \eqref{Y22}.
This requirement gives
\begin{align}
C^{2}_{11}=C^{2}_{22}=C^{2}_{33}=C^{2}_{44},\qquad C^{2}_{55}= C^{2}_{66}= C^{2}_{77}= C^{2}_{88},\qquad
C^{2}_{ij}=0,~{\rm for}~i\ne j.
\end{align}
The traceless condition ($C^2_{ij}\delta^{ij}=0$) implies $C^2_{11}=-C^2_{55}$ while the orthonormality condition gives $C^2_{11}=\frac1{2\sqrt 2}$.

In order to determine the coefficients of the CPOs, $C^{(1)A}_B$, we need to rewrite the scalar spherical harmonics in terms of ${\mathbb C}^4$ coordinates as
\begin{align}\label{Y22b}
Y^{2}=\frac1{L^2}\Big(\sum_{A,B=1}^{2}C^{(1)A}_By_{A}y^{\ast B}+\sum_{A,B=3}^{4} C^{(1)A}_By_Ay^{\ast B}\Big),
\end{align}
where the ${\mathbb C}^4$ coordinates are given by
\begin{align}
y^1=x^1+i x^2,\quad y^2=x^3+i x^4,\quad y^3=x^5+i x^6,
\quad y^4=x^7+i x^8.
\end{align}
Comparing \eqref{Y22} and \eqref{Y22b} we obtain
\begin{align}
&C^{(1)1}_1=\frac{C^{2}_{11}+C^{2}_{22}}2=\frac1{2\sqrt2},\quad C^{(1)2}_2=\frac{C^{2}_{33}+C^{2}_{44}}2=\frac1{2\sqrt2},\nn\\
& C^{(1)3}_3=\frac{C^{2}_{55}+C^{2}_{66}}2=-\frac1{2\sqrt2},\quad C^{(1)4}_4=\frac{C^{2}_{77}+C^{2}_{88}}2=-\frac1{2\sqrt2},\nn\\
&C^{(1)A}_B=0,~{\rm for}~A\ne B.
\end{align}
Hence the CPO of conformal dimension $\Delta=1$ is given by \eqref{Delta1}.

\section{Asymptotic Expansions}\label{AstExp-app}
In this appendix, we display the asymptotic expansion for the warp factors ${\bf g}_{i}(z,\tau)$ in \eqref{dsFG} as well as for the various components of the 4-form field strength ${\bf F}_{pqrs}$ in \eqref{LLMF4} using the FG coordinates defined in \eqref{uxi_trans}.
The expansion can be done to any desirable higher order but we keep only up to $\mu_0^2$ for our purpose.
Applying the ansatze \eqref{uxi_trans2} to the defining functions $Z(\rho,\xi)$ and $V(\rho,\xi)$ in \eqref{ZVfunc}, we obtain the following asymptotic expansions for the warp factors in FG coordinate system,
\begin{align*}
&{\bf g}_1(z,\tau)=\frac{L^2}{4z^2}\Big[1-\mu_0z\frac{\sqrt2(2 C_1^3-3 C_2 C_1+C_3)\tau}{3}\\
&~~~~~-\frac{(\mu_0z)^2}{144}\Big(8 C_1^6-24 C_2 C_1^4-28 C_3 C_1^3+9 \left(5 C_2^2+3 C_4\right) C_1^2-12 C_2 C_3 C_1\\
&~~~~~~~~~~~~~~~~~~~-9 C_2^3+20 C_3^2-27 C_2 C_4\Big)\nn\\
&~~~~~-\frac{(\mu_0z)^2}{144}\Big(28 C_1^6-84 C_2 C_1^4+28 C_3 C_1^3+9 \left(7 C_2^2-15 C_4\right) C_1^2+228 C_2 C_3 C_1\\
&~~~~~~~~~~~~~~~~~~~-135 C_2^3-128 C_3^2+135 C_2 C_4\Big)\tau^2+\cdots\Big],\nn
\end{align*}
\begin{align*}
&{\bf g}_2(z,\tau)=\frac{L^2}{4(1-\tau^2)}\Big[1\\
&~~~~~+\frac{(\mu_0z)^2}{48}\Big(-12 C_1^6+36 C_2 C_1^4+4 C_3 C_1^3-3 \left(13 C_2^2+3 C_4\right) C_1^2+12 C_2 C_3 C_1\\
&~~~~~~~~~~~~~~~~~~~+7 C_2^3-8 C_3^2+9 C_2 C_4\Big)+\cdots\Big],
\end{align*}
\begin{align*}
&{\bf g}_3(z,\tau)=\frac{L^2}{2}(1+\tau)\Big[1+\mu_0z\frac{(2 C_1^3-3 C_2 C_1+C_3)(1+\tau)}{3\sqrt2}\\
&~~~~~+\frac{(\mu_0z)^2}{72}\Big(8 C_1^6-24 C_2 C_1^4-4 C_3 C_1^3+9 \left(3 C_2^2+C_4\right) C_1^2-12 C_2 C_3 C_1\\
&~~~~~~~~~~~~~~~~~~~-3 C_2^3+8 C_3^2-9 C_2 C_4\Big)\nn\\
&~~~~~+\frac{(\mu_0z)^2}{96}\Big(76 C_1^6-228 C_2 C_1^4+28 C_3 C_1^3+3 \left(69 C_2^2-5 C_4\right) C_1^2-12 C_2 C_3 C_1\\
&~~~~~~~~~~~~~~~~~~~-63 C_2^3-8 C_3^2+15 C_2 C_4\Big)\tau\nn\\
&~~~~~+\frac{(\mu_0z)^2}{288}\Big(124 C_1^6-372 C_2 C_1^4+124 C_3 C_1^3+9 \left(31 C_2^2-15 C_4\right) C_1^2+84 C_2 C_3 C_1\\
&~~~~~~~~~~~~~~~~~~~-135 C_2^3-104 C_3^2+135 C_2 C_4\Big)\tau^2+\cdots\Big],\nn
\end{align*}
\begin{align*}
&{\bf g}_4(z,\tau)=\frac{L^2}{2}(1-\tau)\Big[1-\mu_0z\frac{(2 C_1^3-3 C_2 C_1+C_3)(1-\tau)}{3\sqrt2}\\
&~~~~~+\frac{(\mu_0z)^2}{72}\Big(8 C_1^6-24 C_2 C_1^4-4 C_3 C_1^3+9 \left(3 C_2^2+C_4\right) C_1^2-12 C_2 C_3 C_1\\
&~~~~~~~~~~~~~~~~~~~-3 C_2^3+8 C_3^2-9 C_2 C_4\Big)\nn\\
&~~~~~-\frac{(\mu_0z)^2}{96}\Big(76 C_1^6-228 C_2 C_1^4+28 C_3 C_1^3+3 \left(69 C_2^2-5 C_4\right) C_1^2-12 C_2 C_3 C_1\\
&~~~~~~~~~~~~~~~~~~~-63 C_2^3-8 C_3^2+15 C_2 C_4\Big)\tau\nn\\
&~~~~~+\frac{(\mu_0z)^2}{288}\Big(124 C_1^6-372 C_2 C_1^4+124 C_3 C_1^3+9 \left(31 C_2^2-15 C_4\right) C_1^2+84 C_2 C_3 C_1\\
&~~~~~~~~~~~~~~~~~~~-135 C_2^3-104 C_3^2+135 C_2 C_4\Big)\tau^2+\cdots\Big].\nn
\end{align*}
The expansion for the components of the 4-form field strength are also obtained in the same manner
\begin{align*}
&
{\bf F}_{tw_1w_2z}(z,\tau)=-\frac{3L^3}{8z^4}\Big[1-\mu_0z\frac{\sqrt2(2 C_1^3-3 C_2 C_1+C_3)\tau}{3}\\
&~~~~~+\frac{(\mu_0z)^2}{288}\Big(16 C_1^6-48 C_2 C_1^4+28 C_3 C_1^3+27 \left(C_2^2-C_4\right) C_1^2+12 C_2 C_3 C_1\\
&~~~~~~~~~~~~~~~~~~~-15 C_2^3-20 C_3^2+27 C_2 C_4\Big)\nn\\
&~~~~~+\frac{(\mu_0z)^2}{288}\Big(4 C_1^6-12 C_2 C_1^4+4 C_3 C_1^3+9 \left(C_2^2+15 C_4\right) C_1^2-276 C_2 C_3 C_1\\
&~~~~~~~~~~~~~~~~~~~+135 C_2^3+136 C_3^2-135 C_2 C_4\Big)\tau^2+\cdots\Big],\nn\\
&
{\bf F}_{tw_1w_2\tau}(z,\tau)=-\frac{L^3}{8z^3}\Big[\mu_0z\frac{(2 C_1^3-3 C_2 C_1+C_3)}{\sqrt2}\\
&~~~~~-\frac{(\mu_0z)^2}{48}\Big(4 C_1^6-12 C_2 C_1^4+4 C_3 C_1^3+9 \left(C_2^2+15 C_4\right) C_1^2-276 C_2 C_3 C_1\\
&~~~~~~~~~~~~~~~~~~~+135 C_2^3+136 C_3^2-135 C_2 C_4\Big)\tau+\cdots\Big],
\nn\\
&
{\bf F}_{\theta\phi\psi z}(z,\tau)=-\frac{L^3}{8z}(1+\tau)^2\sin\theta\Big[\mu_0z+\frac{(\mu_0z)^2}{3\sqrt2}\Big(2 C_1^3-3 C_2 C_1+C_3\Big)(1+6\tau)+\cdots\Big],\\
&
{\bf F}_{\theta\phi\psi \tau}(z,\tau)=-\frac{L^3}{4}(1+\tau)\sin\theta\Big[\mu_0z+\frac{(\mu_0z)^2}{6\sqrt2}\Big(2 C_1^3-3 C_2 C_1+C_3\Big)(4+9\tau)+\cdots\Big],
\nn\\
&
{\bf F}_{\tilde\theta\tilde\phi\tilde\psi z}(z,\tau)=-\frac{L^3}{8z}(1-\tau)^2\sin\tilde\theta\Big[\mu_0z-\frac{(\mu_0z)^2}{3\sqrt2}\Big(2 C_1^3-3 C_2 C_1+C_3\Big)(1-6\tau)+\cdots\Big],\\
&
{\bf F}_{\tilde\theta\tilde\phi\tilde\psi \tau}(z,\tau)=\frac{L^3}{4}(1-\tau)\sin\tilde\theta\Big[\mu_0z-\frac{(\mu_0z)^2}{6\sqrt2}\Big(2 C_1^3-3 C_2 C_1+C_3\Big)(4-9\tau)+\cdots\Big].
\end{align*}

\section{Proof of \eqref{Identity}}\label{Cpt1}
In this section we prove the relation \eqref{Identity}.
We start from the definition of $C_1$
\begin{align}
C_1=\frac{1}{\sqrt{N}}\sum_{i=1}^{2N_B+1}(-1)^{i+1} m_i,
\end{align}
where we have introduced the integers $m_i=\frac {\tilde x_i}{2\pi l_{\rm P}^3\mu_0}$.
Separating the even and odd terms we rewrite $C_1$ as
\begin{align}
C_1=\frac{1}{\sqrt{N}}\Big[m_1+\sum_{i=1}^{N_B}(m_{2i+1}- m_{2i})\Big]=\frac {1}{2\pi l_{\rm P}^3\mu_0\sqrt{N}}\Big[\tilde x_1+\sum_{i=1}^{N_B}(\tilde x_{2i+1}- \tilde x_{2i})\Big]. 
\end{align}
The term in the square bracket is the position of the Fermi level defined in \eqref{xF}.
Here we choose a coordinate system in which the zero of the $\tilde x$-coordinate is at the position of the Fermi level i.e., $\tilde x_F=0$ and hence $C_1=0$.

With the above choice of coordinate system, we notice that the $\beta_3$ on the right hand side of \eqref{Identity} is equal to $C_3$.
However, for convenience we add $\frac{C_1}{N}$ and write it as
\begin{align}
\beta_3=-\left(\frac{C_1}{N}-C_3\right)=-\frac{1}{N^{3/2}}\sum_{i=1}^{2N_B+1}(-1)^{i+1} (m_i-m_i^3).
\end{align}
Without loss of generality, we choose a droplet in which the Fermi level lies in the black region.
See Fig.1.
Then we can write $\beta_3$ as
\begin{align}
N^{3/2} \beta_3=-\sum_{i=1}^{2j}(-1)^{i+1} (m_i-m_i^3)-\sum_{i=2j+1}^{2N_B+1}(-1)^{i+1} (m_i-m_i^3),
\end{align}
where $m_{2j}$ is the position of the first boundary just below the Fermi level while $m_{2j+1}$ is the position of the first boundary just above the Fermi level.
Since the Fermi level is at zero, we can rewrite the above equation as
\begin{align}
N^{3/2} \beta_3=\sum_{i=1}^{2j}(-1)^{i+1} (|m_i|-|m_i|^3)-\sum_{i=2j+1}^{2N_B+1}(-1)^{i+1} (|m_i|-|m_i|^3).
\end{align}
Using the relation
\begin{align}
\sum_{n=0}^{|m_i|-1}n(n+1)=-\frac13 (|m_i|-|m_i|^3),
\end{align}
we obtain
\begin{align}
\frac{N^{3/2} }{3}\beta_3=\bigg[-\sum_{i=1}^{2j}\sum_{n=0}^{|m_i|-1}+\sum_{i=2j+1}^{2N_B+1}\sum_{n=0}^{|m_i|-1}\bigg](-1)^{i+1}n(n+1).
\end{align}
Noting that for the first double summation, $m_i$'s are negative while they are positive for the second double summation, we write
\begin{align}
\frac{N^{3/2} }{3} \beta_3=\bigg[-\sum_{i=1}^{2j}\sum_{n=0}^{-m_i-1}+\sum_{i=2j+1}^{2N_B+1}\sum_{n=0}^{m_i-1}\bigg](-1)^{i+1}n(n+1).
\end{align}
Next lets expand the summations over $i$, which gives
\begin{align}
\frac{N^{3/2} }{3} \beta_3=&-\bigg[\sum_{n=0}^{-m_{1}-1}-\sum_{n=0}^{-m_{2}-1}+\cdots +\sum_{n=0}^{-m_{2j-1}-1}-\sum_{n=0}^{-m_{2j}-1}\bigg]n(n+1)\nn\\
&+\bigg[\sum_{n=0}^{m_{2j+1}-1}-\sum_{n=0}^{m_{2j+2}-1}+\sum_{n=0}^{m_{2j+3}-1}+\cdots -\sum_{n=0}^{m_{2N_B}-1}+\sum_{n=0}^{m_{2N_B+1}-1}\bigg]n(n+1).
\end{align}
We combine the summations in each square brackets pair by pair, leaving the first term in the second square bracket unpaired, to obtain
\begin{align}
\frac{N^{3/2} }{3} \beta_3=&-\bigg[\sum_{n=-m_{2}}^{-m_{1}-1}+\sum_{n=-m_{4}}^{-m_{3}-1}+\cdots +\sum_{n=-m_{2j}}^{-m_{2j-1}-1}\bigg]n(n+1)\nn\\
&+\bigg[\sum_{n=0}^{m_{2j+1}-1}+\sum_{n=m_{2j+2}}^{m_{2j+3}-1}+\cdots +\sum_{n=m_{2N_B}}^{m_{2N_B+1}-1}\bigg]n(n+1).
\end{align}
We note that the summations in the first square bracket cover the white regions below the Fermi level, while the summations in the second square bracket cover the black regions above the Fermi level.
However, we recall that in the $k=1$ case, the occupation numbers $l_n=1$ for the $n^{th}$ excitation level located in a black region above the Fermi level while $l_n=0$ for the $n^{th}$ excitation level located in a white region above the Fermi level.
Similarly, the occupation numbers $l_n'=1$ for the $n^{th}$ excitation level located in a white region below the Fermi level while $l_n'=0$ for the $n^{th}$ excitation level located in a black region below the Fermi level.
Therefore, we can write the summations in first square bracket as summation over the entire region below the Fermi level by introducing $l_n'$ and also second square bracket as summation over the entire region above the Fermi level by introducing $l_n$. Then we obtain
\begin{align}
\frac{N^{3/2} }{3} \beta_3=\sum_{n=0}^{\infty}\Big[n(n+1)(l_n-l_n')\Big],
\end{align}
which is what we have in \eqref{Identity}.

\end{document}